\documentclass[aps,prc,twocolumn,showpacs,nofootinbib,superscriptaddress]{revtex4-1}
\usepackage{dcolumn}   % needed for some tables
\usepackage{bm}        % for math
\usepackage{multirow}

\usepackage{amsmath}
\usepackage{amsfonts}
\usepackage{amssymb}
\usepackage{makeidx}
\usepackage{graphicx}
\usepackage{anysize}
\usepackage{hyperref}

\usepackage{slashed}

\newenvironment{mymathbox}
{\par\smallskip\centering\begin{lrbox}{0}%
\begin{minipage}[c]{0.8\textwidth}}
{\end{minipage}\end{lrbox}%
\framebox[0.9\textwidth]{\usebox{0}}%
\par\medskip
\ignorespacesafterend}
\newcommand{\bb}{\begin{mymathbox}}
\newcommand{\eb}{\end{mymathbox}}
                
\newcommand{\munit}{\mbox{\boldmath $1\!\!1$}}

\newcommand{\Dslash}{\slashed{D}}

\newcommand{\Kslash}{\slashed{K}}

\newcommand{\Qslash}{\slashed{Q}}

\newcommand{\be}{\begin{equation}}
\newcommand{\ee}{\end{equation}}
\newcommand{\ba}{\begin{eqnarray}}
\newcommand{\ea}{\end{eqnarray}}

\newcommand{\ubarN}{\overline{u}({\bf p}_N,s_N)}
\newcommand{\uu}{u({\bf p},s)}

\newcommand{\nk}{{\bf      k}}
\newcommand{\np}{{\bf      p}}
\newcommand{\nq}{{\bf      q}}

\newcommand{\nPsi}{{\bf \Psi}}
\newcommand{\npsi}{{\bf \npsi}}
\newcommand{\ntau}{{\boldsymbol \tau}}
\newcommand{\vphi}{\vec{\phi}}

\newcommand{\psib}{\overline{\psi}}
\newcommand{\nPsib}{\overline{\nPsi}}

\newcommand{\la}{\langle}
\newcommand{\ra}{\rangle}

\newcommand{\de}{\text{d}}
\newcommand{\pd}{\partial_\mu}
\newcommand{\pa}{\partial^\mu}

\newcommand{\non}{\nonumber}

\newcommand{\bma}{\begin{pmatrix}}
\newcommand{\ema}{\end{pmatrix}}

\newcommand{\Cab}{{\cos\theta_c}}

\usepackage[dvips]{color}

\begin{document}

\title{Electroweak single-pion production off the nucleon: from threshold to high invariant masses}

\author{R.~Gonz\'alez-Jim\'enez}
\email{Raul.GonzalezJimenez@UGent.be}
\affiliation{Department of Physics and Astronomy, Ghent University, Proeftuinstraat 86, B-9000 Gent, Belgium}
\author{N.~Jachowicz}
\affiliation{Department of Physics and Astronomy, Ghent University, Proeftuinstraat 86, B-9000 Gent, Belgium}
\author{K. Niewczas}
\affiliation{Department of Physics and Astronomy, Ghent University, Proeftuinstraat 86, B-9000 Gent, Belgium}
\affiliation{Institute of Theoretical Physics, University of Wroc{\l}aw, pl. M. Borna 9, 50-204 Wroc{\l}aw, Poland}
\author{J.~Nys}
\affiliation{Department of Physics and Astronomy, Ghent University, Proeftuinstraat 86, B-9000 Gent, Belgium}
\author{V.~Pandey}
\affiliation{Center for Neutrino Physics, Virginia Tech, Blacksburg, Virginia 24061, USA}
\author{T.~Van~Cuyck}
\affiliation{Department of Physics and Astronomy, Ghent University, Proeftuinstraat 86, B-9000 Gent, Belgium}
\author{N.~Van~Dessel}
\affiliation{Department of Physics and Astronomy, Ghent University, Proeftuinstraat 86, B-9000 Gent, Belgium}

\date{\today}

\begin{abstract}
\begin{description}
\item[Background] Neutrino-induced single-pion production (SPP) provides an important contribution to neutrino-nucleus interactions, ranging from intermediate to high energies. There exists a good number of {\it low-energy models} in the literature to describe the neutrinoproduction of pions in the region around the Delta resonance. Those models consider only lowest-order interaction terms and, therefore, fail in the high-energy region (pion-nucleon invariant masses, $W\gtrsim2$ GeV).
\item[Purpose] Our goal is to develop a model for electroweak SPP off the nucleon, which is applicable to the entire energy range of interest for present and future accelerator-based neutrino-oscillation experiments. 
\item[Method] We start with the low-energy model of Ref.~\cite{Hernandez07}, which includes resonant contributions and background terms derived from the pion-nucleon Lagrangian of chiral-perturbation theory~\cite{Scherer12}. Then, from the background contributions, we build a high-energy model using a Regge approach. The low- and high-energy models are combined, in a phenomenological way, into a hybrid model.
\item[Results] The Hybrid model is identical to the low-energy model in the low-$W$ region, but, for $W>2$ GeV, it implements the desired high-energy behavior dictated by Regge theory.
We have tested the high-energy model by comparing with one-pion production data from electron and neutrino reactions. 
% The Hybrid model is compared with inclusive (double-differential) electron-proton scattering data and with neutrino one-pion production total cross section data from the ANL and BNL experiments.
The Hybrid model is compared with electron-proton scattering data, with neutrino SPP data and with the predictions of the NuWro Monte Carlo event generator.
\item[Conclusions] Our model is able to provide satisfactory predictions of the electroweak one-pion production cross section from pion threshold to high $W$.
Further investigation and more data are needed to better understand the mechanisms playing a role in the electroweak SPP process in the high-$W$ region, in particular, those involving the axial current contributions.
\end{description}
\end{abstract}


% 13.60.Le : Meson production by photons and leptons
% 13.15.+g : Neutrino interactions with Hadrons
% 25.30.Pt : Neutrinos in nuclear scattering
% 12.15.-y : Weak interactions

\maketitle

% \tableofcontents

\section{Introduction}

Single-pion production constitutes an important contribution to the neutrino-nucleus cross section in the region covered by accelerator-based neutrino-oscillation experiments such as MiniBooNE and T2K~\cite{MiniBooNECC13b,T2KCC14}, with a beam energy of $\varepsilon_\nu\sim0.5-2$ GeV.
These experiments, which use nuclei as target material, select the events of the dominant charged-current (CC) quasielastic (QE) $\nu_\mu n\rightarrow \mu^- p$ channel to reconstruct the neutrino energy.
Single-pion production $\nu_\mu N\rightarrow \mu^- N'\pi$ is an important background in the identification process: if the produced pion is absorbed in the nucleus the signal mimics a QE event in the detector.
These events are subtracted from the QE sample using event generators which base their predictions on theoretical models.
Thus, if such predictions are not accurate, the error is propagated systematically to the reconstructed energy which is subsequently used to obtain the neutrino oscillation parameters.
Weak-neutral current (WNC) $\pi^0$ production, $\nu N\rightarrow \nu N'\pi^0$, is also an important background in $\nu_e$ ($\bar\nu_e$) appearance experiments due to the difficulty to distinguish between a $\pi^0$ and an electron (positron) signal.
Therefore, in order to make precise estimates for the desired oscillation parameters, it is essential to have theoretical models capable of providing reliable predictions for the pion-production process at the vertex level as well as for the pion propagation through the nuclear medium.
In addition, weak pion production offers a unique opportunity to learn about the axial form factors of the nucleon resonances and, in general, about the nucleon axial current.
In this line, the recent sets of neutrino-induced SPP data from MiniBooNE~\cite{MBNCpion010,MBCCpion011,MBCCpionC11} and MINERvA~\cite{MINERvACCpi15,MINERvACCpi015,MINERvApion16}, as well as the inclusive neutrino-nucleus data from T2K~\cite{T2Kinc13} and SciBooNE~\cite{SciBooNE11}, offer an excellent opportunity to test and improve the existing models on pion production in the nuclear medium.

There exists a good number of models in the literature describing the neutrinoproduction of pions in the region around the Delta resonance~\cite{Rein81,Sato03,Amaro05a,Ahmad06,Hernandez07,Buss07,Praet09,Martini09,Serot12,Ivanov16,Rafi16} (see Refs.~\cite{Alvarez-Ruso14,Mosel16,Katori16} for a review), that is, from the pion threshold to $W\approx1.4$ GeV (where $W$ is the $\pi N$-invariant mass).
In spite of the differences between these models, all of them describe the reaction amplitude using lowest order interaction terms, which is appropriate near the pion-production threshold.
At increasing energies, however, higher order contributions must be taken into account to properly model the dynamics of the reaction. It is not surprising, therefore, that low-energy models fail when they are applied in the high-energy region ($W\gtrsim2$ GeV).
It is well-known that the relative importance of SPP in the inclusive cross section gradually decreases for increasing invariant masses due to the contribution of a multitude of new opened channels, such as two-pion production, production of other mesons, deep inelastic scattering (DIS), etc.
In spite of that, it remains highly desirable to have models that are able to make reliable predictions for the SPP channel in the entire energy domain covered by experiments.
This is evident, for instance, in the MINERvA experiment, where the neutrino flux peaks around 4 GeV and extends to 10 GeV, which allows for invariant masses much larger than $1.4$ GeV. 
The same will happen in future accelerator-based neutrino-oscillation experiments, such as NOvA and DUNE~\cite{NOvA07,DUNE16}, where the beam flux covers a similar energy range as the MINERvA experiment.

In this work, we propose to extend the description of neutrino-induced pion production to higher $W$. For that, we use Regge phenomenology~\cite{Collins77,Gribov01}.
As a first step, we focus on neutrino-induced SPP off the nucleon. Work on applying this model to the pion production on nuclei is in progress.
Our starting point is the model first presented by Hern\'andez, Nieves and Valverde in Ref.~\cite{Hernandez07}, and later improved in Refs.~\cite{Hernandez13,Alvarez-Ruso16}. 
This model provides a microscopic description of the SPP by working at the amplitude level. 
The pion-production mechanism includes, in addition to the dominant Delta-resonance decay, the background contributions deduced from the chiral-perturbation theory Lagrangian of the pion-nucleon system~\cite{Scherer12} (ChPT background or, simply, background terms in what follows). 
Later on, the $D_{13}(1520)$ resonance was included in the model in order to improve the comparison with MiniBooNE data~\cite{Hernandez13}.
Recently, the model was further improved by incorporating the relative phases between the background diagrams and the dominant partial wave of the Delta pole, which partially restores unitarity (see Ref.~\cite{Alvarez-Ruso16} for details).
This model has been used by several groups with notable success in reproducing electron- and neutrino-induced SPP data in the region $W\lesssim1.4$ GeV~\cite{Lalakulich10,Sobczyk13,Rafi16}.

Regge phenomenology is a well-tested method that allows one to make predictions in the high-energy domain ($W>2$ GeV) without having to resort to partonic degrees of freedom. It has proven successful in modeling purely hadronic interactions~\cite{Collins77}, as well as photo- and electroproduction on nucleons~\cite{Guidal97,Kaskulov10a}. 
For increasing $W$, the number of overlapping nucleon resonances grows and the typical peaked resonance structures observed at low energies tends to disappear. The main idea that drives Regge theory is that at high energies the dynamics can be understood in terms of cross-channel contributions only. 
This observation is generally explained by a duality argument, which states that the amplitude obtained by summing over all $s$-channel resonances is, on average, equivalent to the amplitude obtained by summing over all $t$-channel meson-exchange amplitudes (or Regge poles). 
Thus, only background terms are normally considered in Regge-based models. Care must be taken, since a truncated set of cross-channel contributions continued into the direct channel generally results in a disastrous overshoot of the cross section. Therefore, in Regge theory, one analytically continues a summation of an infinite number of cross-channel partial waves into the direct channel.

We propose to {\it reggeize} the ChPT-background contributions by following the method originally presented by Levy, Majerotto, and Read~\cite{Levy73}, and further developed by Guidal, Laget, and Vanderhaeghen~\cite{Guidal97} for charged-meson photoproduction at high $W$ and forward scattering angles of the pion. The method consists in replacing the propagator of the $t$-channel meson-exchange diagrams by the corresponding Regge propagators. In this way, each ($t$-channel) Regge amplitude includes an entire family of hadrons. 
The great success achieved by this model~\cite{Guidal97} in reproducing high-energy forward-scattering photoproduction data, motivated its extension to other reactions such as electroproduction~\cite{Vanderhaeghen98,Laget04,Kaskulov08,Kaskulov10a,Vrancx14a,Vrancx14b}, which is of major interest to us. In the framework of Ref.~\cite{Guidal97}, the $s$- and $u$-channel nucleon poles are considered along with the $t$-channel pion-exchanged diagram in order to preserve vector-current conservation. 
While this approach allowed to understand the behavior of the longitudinal component of the electroproduction cross section, it did not provide a satisfactory description of the transverse one. This was remedied by Kaskulov and Mosel in Ref.~\cite{Kaskulov10a}, by taking into account the fact that the nucleon in the $s$- and $u$-channel Born diagrams may be highly off its mass shell. 
We will use this model as a guide to reggeize the ChPT-background terms. 
Based on the mentioned previous works, it is straightforward to reggeize the vector current of the ChPT background, the novelty of this work resides in the reggeization of the axial current, which allows us to make predictions for neutrino interaction.

The outline of this article is as follows. 
In Sec.~\ref{Kinematic}, we present the kinematic relations and the cross section formula.
In Sec.~\ref{Low-energy-model}, we briefly summarize the low-energy model.
In Sec.~\ref{High-energy-model}, we discuss, in detail, the procedure to build the high-energy model by reggeizing the background contributions of the low-energy model.
In Sec.~\ref{Hybrid-model}, we propose a phenomenological way of combining the low- and high-energy models into a hybrid model.
Our conclusions are summarized in Sec.~\ref{Conclusions}.
We have tried to make this article as self-contained as possible. 
For this reason, we have included in Appendix~\ref{ChPT-expressions} the derivation of the axial and vector currents of the pion-nucleon system in the ChPT framework.  
Finally, in Appendix~\ref{nucleon-resonances} we have included the formulas and parameters related to the nucleon resonances that are considered in this work.

\section{Kinematics and cross section} \label{Kinematic}

\begin{figure}[htbp]
  \centering  
      \includegraphics[width=.5\textwidth,angle=0]{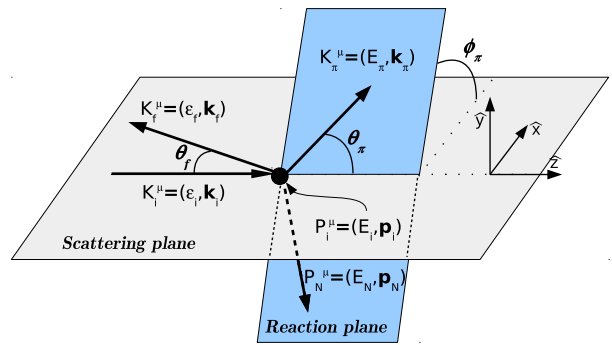}
  \caption{One-pion production on a free nucleon in the laboratory system. 
  An incoming lepton $K_i^\mu$ along the $\hat z$ direction is scattered by a free nucleon at rest $P_i^\mu=(M,\bf{0})$. 
  The final state is given by a lepton $K_f^\mu$, a nucleon $P_N^\mu$, and a pion $K_\pi^\mu$.}
  \label{fig:kin-free}
\end{figure}

We aim at modeling the electroweak one-pion production on free nucleons. Electroweak refers to the fact that the same model is used to describe electron-induced one-pion production, mediated by the exchange of a virtual photon (electromagnetic interaction, EM), as well as neutrino-induced one-pion production, mediated by the exchange of W$^\pm$ boson (CC interaction) or Z boson (WNC interaction). 
The process is depicted in Fig.~\ref{fig:kin-free} along with the notation employed for the kinematic variables. 
All four-vectors involved in the $2 \to 3$ process (Fig.~\ref{fig:kin-free}) are completely determined by six independent variables. We therefore introduce the ``lab variables'' $\varepsilon_i$, $\varepsilon_f$, $\theta_f$, $\phi_f$, $\theta_\pi$, and $\phi_\pi$~\footnote{Since the process is symmetric under rotation over the angle $\phi_f$, we can set $\phi_f=0$. Thus, the actual number of independent variables needed to characterize the process reduces to five.}.

The most exclusive cross section for the process depicted in Fig.~\ref{fig:kin-free} is given by
\ba
  \frac{\de^5\sigma}{\de \varepsilon_f\de\Omega_f\de\Omega_\pi} = \frac{{\cal F}}{(2\pi)^5 } 
      \frac{\varepsilon_f k_f k_\pi E_\pi }{f_{rec}} l_{\mu\nu}h^{\mu\nu}
\ea
with
\ba
  f_{rec} = \left|1 + \frac{E_\pi}{E_N k_\pi^2}\ \nk_\pi\cdot(\nk_\pi-\nq) \right|\,,
\ea
the nucleon recoil factor.
The factor ${\cal F}$ depends on the particular process under study and includes the boson propagator as well as the coupling constants at the leptonic vertex:
\ba
  {\cal F}_{EM} = \frac{e^2}{Q^4}\,,\,\,\,\,\,   
  {\cal F}_{CC} = \frac{g^2}{8M_W^4}\,,\,\non\\ 
  {\cal F}_{WNC} = \frac{g^2}{16M_Z^4 \cos^2\theta_W}\,.
\ea
We have introduced $Q^2 = -(K_i-K_f)^2 = \nq^2-\omega^2>0$, with $\omega$ and $\nq$ the energy and momentum transfer in the lab system.
We define the dimensionless lepton tensor $l_{\mu\nu}$ as 
\ba
  l_{\mu\nu}^{EM} &=& \frac{1}{2\varepsilon_i\varepsilon_f} s_{\mu\nu}\,,\\ 
  l_{\mu\nu}^{CC} &=& l_{\mu\nu}^{WNC} = \frac{2}{\varepsilon_i\varepsilon_f} \left( s_{\mu\nu} - iha_{\mu\nu} \right)\non
\ea
with $h=-1(+1)$ for neutrinos (antineutrinos). 
The lepton tensor has been decomposed in terms of the symmetric ($s_{\mu\nu}$) and antisymmetric ($a_{\mu\nu}$) tensors:
\ba
  s_{\mu\nu} &=& K_{i,\mu}K_{f,\nu} + K_{f,\mu}K_{i,\nu} - g_{\mu\nu}K_{i}\cdot K_{f},\non\\
  a_{\mu\nu} &=& \epsilon_{\mu\nu\alpha\beta} K^\alpha_{i}K^\beta_{f}\label{eq:amunu}
\ea
with $\epsilon_{0123}=+1$. 

The hadronic tensor is given by
\ba
  h^{\mu\nu} = \frac{M}{2E_\pi E_N}\overline{\sum_{s_N,s}} \la J^\mu \ra^\dagger \la J^\nu\ra\,,
\ea
where $\overline{\sum}_{s_N,s}$ implies a sum over the spin of the final nucleon ($s_N$) and an average over the spin of the initial nucleon ($s$). The hadronic current $\la J^\mu\ra$ is evaluated between initial and final states and, in general, depends on all independent variables of the process. For simplicity, in what follows we will simply write $J^\mu$.
It reads
\ba
  J^\mu = {\cal C}\ \ubarN {\cal O}_{1\pi}^\mu \uu \,,\label{eq:Jhad}
\ea
where ${\cal O}_{1\pi}^\mu$ represents the operator which induces the transition between the initial one-nucleon state and the final one-nucleon one-pion state. 
$\ubarN$ and $\uu$ are free Dirac spinors describing the scattered and initial nucleon, respectively.
${\cal C}$ is the electroweak coupling constant of the hadronic current:
\begin{flalign}
  {\cal C}_{CC} = \frac{ig\Cab}{2\sqrt2},\ {\cal C}_{WNC} = \frac{ig}{2\cos\theta_W},\
  {\cal C}_{EM} = ie. \label{eq:C-const}
\end{flalign}

\begin{figure}[htbp]
  \centering  
      \includegraphics[width=.4\textwidth,angle=0]{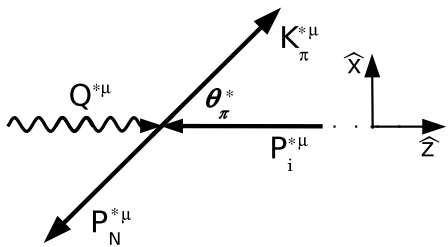}
  \caption{Hadronic $2 \to 2$ subprocess in the center of mass system.  }
  \label{fig:cms}
\end{figure} 

It is convenient to introduce variables which intuitively describe the hadronic $2 \to 2$ subprocess (Fig.~\ref{fig:cms}). 
In the hadronic center of mass system (cms)~\footnote{The variables in the cms are denoted by the $^*$ index.}, where $\np^*_i+\nq^*=0$ and $\nq^*$ is defined along the $\hat z$ direction (see Fig.~\ref{fig:cms}), only three independent variables are required to describe the kinematics of this subsystem. Therefore, we introduce the set of variables $s$, $t$ and $Q^2$, with $s$ and $t$ the usual Mandelstam variables. To facilitate discussions about the dynamics of the system, we determine the kinematic regime of the hadronic subprocess related to a given lab configuration.

\begin{figure}[htbp]
  \centering  
      \includegraphics[width=.4\textwidth,angle=270]{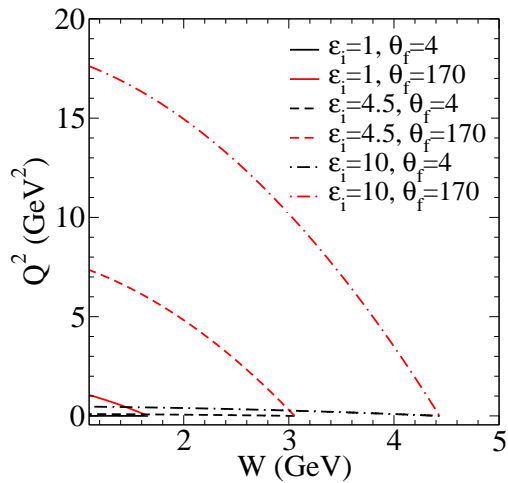}
  \caption{(Color online) $Q^2$ as a function of the invariant mass $W$ for different kinematics (see text for details). 
  The curves obey Eq.~\ref{eq:W2-Q2} in the case $m_f\rightarrow0$. }
  \label{fig:Q2-vs-W}
\end{figure}

In general, the invariant mass $W$ is given by $W=\sqrt s$, with 
\ba
  s = (P_i + Q)^2 = M^2 + 2M\omega - Q^2\,.\label{eq:W2-Q2}
\ea 
Some of the results presented in the next sections are computed in two extreme kinematic scenarios: backward scattering ($\theta_f\rightarrow 180$ deg), which correspond to high $Q^2$, and forward scattering ($\theta_f\rightarrow0$ deg), corresponding to $Q^2\approx0$.   
The idea is that given the results at these kinematics, it is possible to intuit or extrapolate the behavior at any other intermediate situation. 
In Fig.~\ref{fig:Q2-vs-W}, we show the relation between the invariant mass $W$ and $Q^2$, according to Eq.~\ref{eq:W2-Q2}: 
we fix ($\theta_f$, $\varepsilon_i$) and vary $\varepsilon_f$. 
Thus, for a fixed value of $\varepsilon_i$, any curve corresponding to an intermediate scattering angle $\theta_f$ will fall in the region between the two extreme curves.

\begin{figure}[htbp]
  \centering  
      \includegraphics[width=.4\textwidth,angle=0]{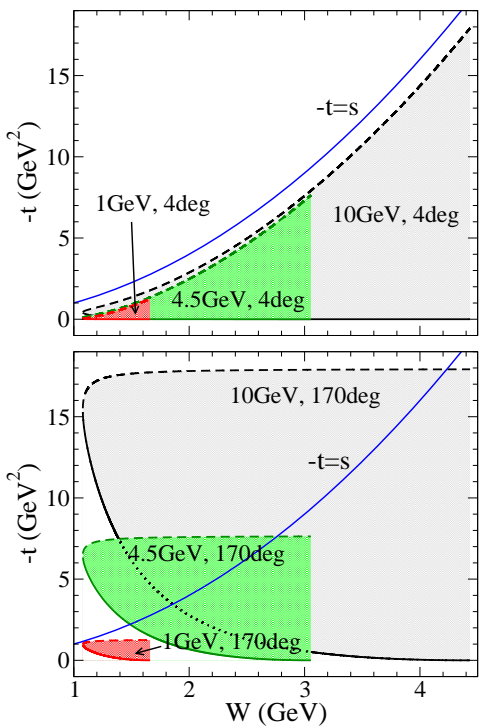}
  \caption{(Color online) Each colored area represents the $-t$ region allowed by energy-momentum conservation for fixed $(\varepsilon_i,\theta_f)$ values: $\varepsilon_i=$ 1 (red), 4.5 (green) and 10 GeV (gray).  
  The top panel corresponds to $\theta_f=4$ deg and the bottom panel to $\theta_f=170$ deg.
  The $-t$-maximum (dashed) and $-t$-minimum (solid) lines are solutions of Eq.~\ref{eq:textreme} for $\cos\theta_\pi^*=-1$ and $\cos\theta_\pi^*=1$, respectively. 
  The blue solid line is the curve corresponding to $-t=W^2$.}
  \label{fig:tregion}
\end{figure}

In the cms (Fig.~\ref{fig:cms}), the Mandelstam variable $t=(Q-K_\pi)^2$ can be related to the cms-pion scattering angle $\theta^*_\pi$  by 
\ba
 t = m_\pi^2 - Q^2 - 2 ( \omega^* E_\pi^* -q^* k_\pi^*\cos\theta^*_\pi )\,, \label{eq:textreme}
\ea
where
\begin{align}
  \omega^* = \frac{s-M^2-Q^2}{2W}\,,\,\,
  E_\pi^* = \frac{s+m_\pi^2-M^2}{2W}\,.\label{eq:Epi*}
\end{align}
In the case $\cos\theta_\pi^*=-1$ and $\cos\theta_\pi^*=1$, Eq.~\ref{eq:textreme} provides the maximum and minimum $t$ values allowed by energy-momentum conservation. 
This is shown in Fig.~\ref{fig:tregion} for the same set of kinematics as in Fig.~\ref{fig:Q2-vs-W}. 
Fixing ($\theta_f$, $\varepsilon_i$, $\cos\theta_\pi^*=\pm1$) and varying $\varepsilon_f$ we generate the solid and dashed lines, which correspond to the extreme values of $t$. 
The shadowed area between these curves represents the allowed kinematic region.
The blue solid line is the parabola corresponding to $-t = s$. 
We set the limit of applicability of our high-energy model (based on Regge-trajectory exchanges) to the region below this line and $W>2$ GeV (see Sec.~\ref{High-energy-model}).

\section{Low energy model: Resonances and ChPT background}\label{Low-energy-model}

\begin{figure}[htbp]
  \centering  
      \includegraphics[width=.15\textwidth,angle=0]{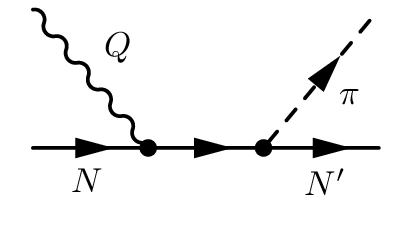}
      \includegraphics[width=.15\textwidth,angle=0]{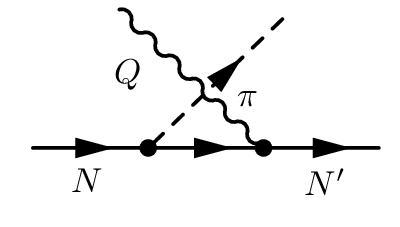}\\
      \includegraphics[width=.15\textwidth,angle=0]{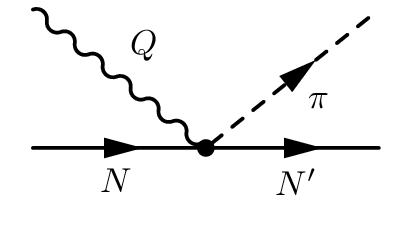}
      \includegraphics[width=.15\textwidth,angle=0]{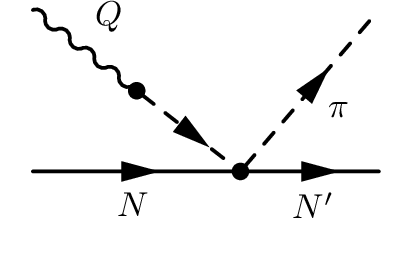}
      \includegraphics[width=.15\textwidth,angle=0]{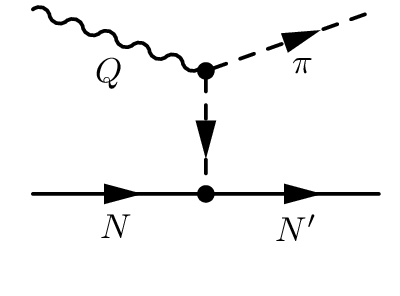}
  \caption{ChPT-background contributions (from left to right and top to bottom): $s$ channel (nucleon pole, $N P$), $u$ channel (cross-nucleon pole, $C N P$), contact term ($CT$), pion pole ($PP$), and $t$ channel (pion-in-flight term, $PF$).}
  \label{fig:nres-diagrams}
\end{figure}

\begin{figure}[htbp]
  \centering  
      \includegraphics[width=.15\textwidth,angle=0]{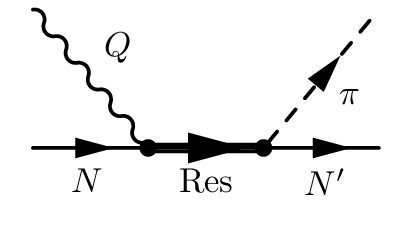}
      \includegraphics[width=.15\textwidth,angle=0]{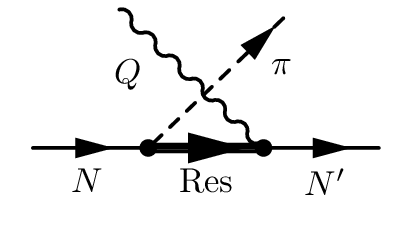}
  \caption{$s$- and $u$-channel diagrams for the nucleon resonances.}
  \label{fig:res-diagrams}
\end{figure}

In this section, we describe the low-energy model. It contains the $s$- and $u$-channel diagrams of the $P_{33}(1232)$ (Delta) and $D_{13}(1520)$ resonances and the background terms from the ChPT $\pi N$-Lagrangian (Appendix~\ref{ChPT-expressions}), as presented in Refs.~\cite{Hernandez07,Hernandez13}.
In addition, we also consider the $s$- and $u$-channel contributions from the spin-$1/2$ resonances $S_{11}(1535)$ and $P_{11}(1440)$.
The corresponding Feynman diagrams are shown in Figs.~\ref{fig:nres-diagrams} and \ref{fig:res-diagrams}. 
In the following, we summarize the expressions for the hadronic current operators (see Eq.~\ref{eq:Jhad}) for each contribution.

The hadronic current operators for the background terms of Fig.~\ref{fig:nres-diagrams} are:
\ba
 {\cal O}^\mu_{NP} = i{\cal I}\,\frac{-g_A}{\sqrt2f_\pi} \Kslash_\pi\gamma^5 \frac{\Kslash_s + M}{s-M^2} \hat\Gamma^\mu_{QNN}\,, \label{eq:ONP}
\ea
with $K_s^\mu=P^\mu+Q^\mu$, and $\hat\Gamma^\mu_{QNN}(Q^\mu)$ given in Eqs.~\ref{eq:GgNN}, \ref{eq:GWNN} and \ref{eq:GZNN},
\ba
 {\cal O}^\mu_{CNP} = i{\cal I}\, \frac{-g_A}{\sqrt2f_\pi} 
	  \hat\Gamma^\mu_{QNN} \frac{\Kslash_u + M}{u-M^2} \Kslash_\pi\gamma^5\,, \label{eq:OCNP} 
\ea
with $K_u^\mu=P^\mu-K_\pi^\mu$ and $u=K_u^2$,
\ba
 {\cal O}^\mu_{PF} = i{\cal I}\, F_{PF}(Q^2) \frac{-g_A}{\sqrt2f_\pi}
	  \frac{K_\pi^\mu-K_t^\mu}{t-m_\pi^2} \Kslash_t\gamma^5\,, \label{eq:OPF}
\ea
with $K_t^\mu=Q^\mu-K_\pi^\mu$, 
\begin{align}
 {\cal O}^\mu_{PP} = i{\cal I}\, F_{\rho}(t) \frac{-1}{\sqrt2f_\pi} 
	  \frac{Q^\mu}{Q^2-m_\pi^2}  \frac{\left(\Qslash + \Kslash_\pi \right)}{2}\,,\label{eq:OPP}
\end{align}
${\cal O}^\mu_{CT} = {\cal O}^\mu_{CTv} + {\cal O}^\mu_{CTa}$ with the axial ($CTa$) and a vector ($CTv$) contributions given by 
\ba 
 {\cal O}^\mu_{CTa} &=& i{\cal I}\, F_{\rho}(t) \frac{1}{\sqrt2f_\pi} \gamma^\mu\,, \label{eq:OCTA}\\
 {\cal O}^\mu_{CTv} &=& i{\cal I}\, F_{CT}(Q^2) \frac{-g_A}{\sqrt2f_\pi} \gamma^\mu\gamma^5\,. \label{eq:OCTV}
\ea
${\cal I}$ is the isospin coefficient of each diagram (see Tables~\ref{Iso-coef-CC} and \ref{Iso-coef-EM}).

We have introduced the nucleon form factors ($F_{1,2}^{p,n}$ for neutral-current interactions and $F_{1,2}^V$ for CC interaction) in the $NP$ and $CNP$ amplitudes (see Eqs.~\ref{eq:GgNN}, \ref{eq:GWNN} and \ref{eq:GZNN}). Therefore, to respect conservation of vector current (CVC) we have included the isovector nucleon form factors, $F_{1,2}^V$, in the $PF$ and $CTv$ amplitudes~\cite{Hernandez07}: 
\ba   
  F_{PF}(Q^2) &=& F_{CT}(Q^2) = F_1^V(Q^2)\non\\ &=& F_1^p(Q^2) - F_1^n(Q^2)\,.
\ea
The form factor $F_\rho(t) = m_\rho^2/(m_\rho^2-t)$, with $m_\rho=775.8$ MeV, has been introduced in the $PP$ term to account for the $\rho$-dominance of the $\pi\pi NN$ coupling~\cite{Towner92,Hernandez07}. 
Consequently, to preserve partial conservation of axial current (PCAC) the same form factor was inserted in the $CTa$ amplitude.

In the case of the WNC interaction, one has to replace the isovector vector form factor by the corresponding WNC one, i.e. $F_1^V \longmapsto \widetilde F_1^V$, with
\ba
 \widetilde F_1^V = \widetilde F_1^p - \widetilde F_1^n = (1-2\sin^2\theta_W)F_1^V\,.
\ea

For nucleon resonances with spin $S=3/2$ (quoted as $R_3$) the direct and cross terms are
\ba
  {\cal O}^\mu_{R_3 P} = i{\cal I}\, \Gamma_{R_3\pi N}^\alpha S_{R_3,\alpha\beta} \Gamma_{QR_3 N}^{\beta\mu}\label{eq:ORP-3/2}\,,\\
  {\cal O}^\mu_{CR_3 P} = i{\cal I}\, \overline{\Gamma}_{QR_3 N}^{\alpha\mu} S_{R_3,\alpha\beta} \Gamma_{R_3\pi N}^\beta\,. \label{eq:OCRP-3/2}
\ea
In the $CR_3 P$ operator (Eq.~\ref{eq:OCRP-3/2}) we introduced the electroweak vertex function
$\overline{\Gamma}_{QR_3 N}^{\beta\mu}(P_N^\mu,Q^\mu) = \gamma^0\left[\Gamma_{QR_3 N}^{\beta\mu}(P_N^\mu,-Q^\mu)\right]^\dagger\gamma^0$.
Finally, for the nucleon resonances with $S=1/2$ ($R_1$) one has
\ba
  {\cal O}^\mu_{R_1 P} = i{\cal I}\, \Gamma_{R_1\pi N} S_{R_1} \Gamma_{QR_1 N}^\mu\,,\label{eq:ORP-1/2} \\ 
  {\cal O}^\mu_{CR_1 P} = i{\cal I}\, \Gamma_{QR_1 N}^\mu S_{R_1} \Gamma_{R_1\pi N} \label{eq:OCRP-1/2}\,.
\ea
The explicit expressions for the strong $R\pi N$ and electroweak $QR N$ vertices, and the resonance propagators ($S_{R}$) are given in Appendix~\ref{nucleon-resonances}.
The isospin coefficients of the Delta resonance are shown in Tables~\ref{Iso-coef-CC} and \ref{Iso-coef-EM}.
The isospin coefficients of the $S_{11}(1535)$ and $P_{11}(1440)$ resonances coincide with those of the $NP$ and $CNP$ terms.

The vector form factors that enter in the electroweak $QR N$ vertices were fitted to reproduce pion photo- and electroproduction data given in terms of helicity amplitudes for the EM current.  
The information on the axial form factors is limited to restrictions provided by the PCAC hypothesis and the BNL and ANL experimental data. 
In this work, we use the parametrization of the vector form factors from Ref.~\cite{Lalakulich06} (for the $P_{33}$, $D_{13}$, and $S_{11}$) and Ref.~\cite{Hernandez08} (for the $P_{11}$).
For the axial form factors we use the parametrizations presented in Ref.~\cite{Alvarez-Ruso16} (for the $P_{33}$), Ref.~\cite{Lalakulich06} (for the $S_{11}$ and $D_{13}$), and Ref.~\cite{Hernandez08} (for the $P_{11}$). 
The explicit expressions for the form factors are given in Appendix~\ref{nucleon-resonances}.

\begin{table}[htbp]
\centering
\begin{tabular}{c|c c c c c}
   Channel 			& $\Delta P$\, 	& $C\Delta P$\, & $NP$\, 	& $CNP$\, 	& Others\\
\hline
 $p \rightarrow \pi^+ + p$ 	& $\sqrt{3/2}$ 	& $\sqrt{1/6}$ 	& 0 		& 1 		& 1\\ 
 $n \rightarrow \pi^0 + p$ 	& $-\sqrt{1/3}$ & $\sqrt{1/3}$ 	& $\sqrt{1/2}$ 	& $-\sqrt{1/2}$ & $-\sqrt{2}$\\  
 $n \rightarrow \pi^+ + n$ 	& $\sqrt{1/6}$ 	& $\sqrt{3/2}$ 	& 1 		& 0 		& $-1$\\ 
\hline 
 $n \rightarrow \pi^- + n$ 	& $\sqrt{3/2}$ 	& $\sqrt{1/6}$ 	& 0 		& 1 		& 1\\
 $p \rightarrow \pi^0 + n$ 	& $\sqrt{1/3}$ 	& $-\sqrt{1/3}$ & $-\sqrt{1/2}$ &  $\sqrt{1/2}$ & $\sqrt{2}$ \\ 
 $p \rightarrow \pi^- + p$ 	& $\sqrt{1/6}$ 	& $\sqrt{3/2}$ 	& 1 		& 0 		& $-1$\\ 
\hline
\end{tabular}
\caption{Isospin coefficients (${\cal I}$) for the different reaction channels in the case of CC interactions.
The first three rows correspond to neutrino-induced reactions. The second three rows correspond to their antineutrino counterparts.}
\label{Iso-coef-CC}
\end{table}
 
\begin{table}[htbp]
\centering
\begin{tabular}{c|c c c c c}
 Channel 		   & $\Delta P$    & $C\Delta P$   & $NP$ 	   & $CNP$ 	   & Others\\
\hline
 $p \rightarrow \pi^0 + p$ & $\sqrt{1/3}$  & $\sqrt{1/3}$  & $\sqrt{1/2}$  & $\sqrt{1/2}$  & 0\\ 
 $p \rightarrow \pi^+ + n$ & $-\sqrt{1/6}$ & $\sqrt{1/6}$  & 1 		   & 1 		   & $-1$\\ 
 $n \rightarrow \pi^- + p$ & $\sqrt{1/6}$  & $-\sqrt{1/6}$ & 1 		   & 1 		   & 1\\ 
 $n \rightarrow \pi^0 + n$ & $\sqrt{1/3}$  & $\sqrt{1/3}$  & $-\sqrt{1/2}$ & $-\sqrt{1/2}$ & 0\\ 
\hline
\end{tabular}
\caption{Isospin coefficients (${\cal I}$) for the different reaction channels in the case of EM and WNC interactions.}
\label{Iso-coef-EM}
\end{table}

The current operator that enters in the hadronic current of Eq.~\ref{eq:Jhad} is given by 
${\cal O}_{1\pi}^\mu = {\cal O}_{ChPT}^\mu + \sum_{R}{\cal O}_{R}^\mu$ with 
\ba
  {\cal O}_{ChPT}^\mu &=& {\cal O}^\mu_{NP} + {\cal O}^\mu_{CNP} + {\cal O}^\mu_{PF}\non\\ &+& {\cal O}^\mu_{CTv} + {\cal O}^\mu_{CTa}  + {\cal O}^\mu_{PP}\,,\label{eq:O-ChPT}
\ea
and 
\ba
  {\cal O}_{R}^\mu = {\cal O}^\mu_{R P} + {\cal O}^\mu_{CR P}.\label{eq:O-HR}
\ea
${\cal O}_{R}^\mu$ is the current operator of a resonance $R$.

We have taken into account the relative phases between the ChPT background and the $\Delta P$ contribution making use of the parameterization of the Olsson phases presented in Ref.~\cite{Alvarez-Ruso16}. In this way, unitarity is partially restored (see Ref.~\cite{Alvarez-Ruso16} for details).
In principle, the same should be done for the other higher mass resonances, however, those phases are unknown within this model.
Although not shown here, we have compared all the results presented in this work with the results in the case of adding the higher mass resonances incoherently (i.e., avoiding interferences). We concluded that, given the large uncertainties from others sources such as resonance form factors, the differences between the two approaches are not significant. 
In this regard, the dynamical coupled-channels model for neutrino-induced meson production presented in Refs.~\cite{Sato03,Kamano12,Nakamura15} is worth mentioning. 
To our knowledge, this is the only model that fully controls the interferences between resonant and non-resonant amplitudes.

The low-energy model described above, in particular the model containing the ChPT-background terms and the $s$- and $u$-channel Delta resonance, has been extensively tested versus photon-, electron- and neutrino-induced one-pion production data~\cite{Hernandez07,Lalakulich10,Hernandez13,Sobczyk13,Hilt13,ZmudaPhD,Rafi16}. 
The agreement with data is generally good in the region $W\lesssim1.3-1.4$ GeV where the chiral expansion is expected to be valid and the pathological high-energy behavior of the tree-level diagrams still does not manifest itself.
The incorporation of higher mass resonances does not significantly change the results in this $W$ region so we do not repeat such systematic comparisons with data here.
Instead, in Sec.~\ref{Hybrid-model} we present some results in which this model (referred as LEM in that section) is compared with electron and neutrino cross section data.

\section{High energy model: Reggeizing the ChPT background}\label{High-energy-model}

It is well-known that low-energy models, like the one described above, are not reliable at high invariant masses ($W>1.4$ GeV), due to the fact that only the lowest-order contributions are considered. For increasing energies, one needs to consider higher-order contributions to the amplitude, a procedure that soon becomes intractable and cumbersome. The starting point of Regge theory is an infinite summation over all partial waves in the $t$-channel amplitude. This can be done by reformulating this infinite series in terms of a contour integral in the complex angular momentum plane. A Regge pole then corresponds to a pole in the complex angular momentum plane, which physically represents a whole family of $t$-channel contributions. 
This formalism allows one to incorporate an infinite number of contributions to the scattering amplitude in an efficient way.
Regge theory provides one with the $s$-dependence of the hadronic amplitude at high energies. However, it does not predict the $t$-dependence of the residues~\footnote{Obviously, as for isobar models, Regge theory does not predict the $Q^2$ dependence of the cross section either, since it starts from hadronic degrees of freedom in the cross channel.}. In spite of that, as shown by Guidal, Laget and Vanderhaeghen~\cite{Guidal97}, a good approximation for the $t$-dependence of the residues at forward $\theta_\pi^*$ scattering angles (corresponding to small $-t$ values) can be made from the $t$-dependence of background contributions in low-energy models. 
This is inspired by the fact that the physical region of the direct channel is not too far from the nearest pole of the Regge trajectory. 

In this section, we focus on the modeling of electroweak pion production in the high $W$ and forward $\theta_\pi^*$ scattering region, equivalently, $-t/s<<1$.
To this end, we reggeize the ChPT-background model introduced in the previous section following the method proposed in Ref.~\cite{Guidal97} for photoproduction. This method boils down to replacing the propagator of the $t$-channel meson-exchange diagrams with the corresponding Regge propagators.

At backward $\theta_\pi^*$ scattering angles (corresponding to large $-t$ values), the pion pole is too far from the actual kinematics of the reaction to cause any significant effect in the physical amplitude. Therefore, the $t$-dependence of the amplitude cannot be modeled by $t$-channel meson-exchange diagrams. 
Instead, at backward $\theta_\pi^*$, small $u$ values are approached, which places the $u$-channel nucleon pole closer to the kinematics of the reaction. Consequently, the amplitude should be modeled by the exchange of baryon trajectories in the $u$ channel.
This modeling of the backward reaction is out of the scope of the present work.
Hence, we expect to underestimate the cross section in the backward region. 
This should not be too big a problem as, at high $W$, the magnitude of the cross section at forward $\theta_\pi^*$ scattering is generally orders of magnitude larger than the cross section at intermediate or backward scattering angles (this is shown, for instance, in Fig.~1 of Ref.~\cite{Guidal97}).
It is expected, therefore, that $t$-channel meson-exchange models, as the one presented here, will also provide good predictions of the $\theta_\pi^*$-integrated cross section.

First, we reggeize the EM current and compare our predictions with exclusive $p(e,e'\pi^+)n$ and $n(e,e'\pi^-)p$ experimental data. 
Then, using this model as a benchmark, the formalism is extended to CC and WNC neutrino-induced SPP.

Other studies in which Regge phenomenology is used for describing neutrino-induced SPP were previously presented in Refs.~\cite{Pais70,Gershtein80,Komachenko87,Rein86,Allen86}.

\subsection{Electroproduction of charge pions}\label{regge-electrons}

The building blocks of the Regge-based models for $\pi^{\pm}$ production presented in Refs.~\cite{Guidal97,Vanderhaeghen98,Kaskulov10a,Vrancx14a} are the pion-exchange $t$-channel amplitude and the $s$- and $u$-channel amplitudes with pseudoscalar coupling, which are included to restore CVC.
In Ref.~\cite{Kaskulov10a,Vrancx14a}, only the CVC-restoring component of the $s$($u$)-channel amplitude are kept (i.e. only the ``electric'' coupling is included, which is equivalent to setting $F_1^n=F_2^{p,n}=0$ and $F_1^p\neq0$). 
Under this assumption, the photon does not couple to the neutron and the $p(e,e'\pi^+)n$ amplitude reduces to 
\ba
J^\mu \sim \ubarN \left( {\cal O}_{PF} + {\cal O}^{ps}_{NPv} \right) \uu\,,\label{eq:JpsNP}
\ea
while the $n(e,e'\pi^-)p$ amplitude is given by
\ba
J^\mu\sim\ubarN \left( {\cal O}_{PF} + {\cal O}^{ps}_{CNPv} \right) \uu\,.\label{eq:JpsCNP}
\ea
Here, ${\cal O}^{ps}_{NPv}$ and ${\cal O}^{ps}_{CNPv}$ represent the EM $s$- and $u$-channel Born amplitudes with pseudoscalar-$\pi NN$ coupling, while Eqs.~\ref{eq:ONP} and \ref{eq:OCNP} are the analogous ones with pseudovector coupling. The index $v$ indicates vector current contribution.

Once the $t$-dependence of the amplitude is set up by the tree-level Feynman diagrams, the model is reggeized by replacing the pion propagator with a Regge trajectory: $(t-m_\pi^2)^{-1}\rightarrow {\cal P}_\pi[\alpha_\pi(t)]$, where $\alpha_\pi(t)$ is determined by the experimentally observed spin-mass relation of the pion and its excitation spectrum.
Since the $t$-channel amplitude is multiplied by the factor ${\cal P}_\pi[\alpha_\pi(t)](t-m_\pi^2)$, CVC requires that also the $s$($u$)-channel diagrams are multiplied by the same factor. 

In Ref.~\cite{Guidal97}, this gauge-invariant electric Regge model was applied to meson photoproduction ($Q^2=0$) and, therefore, a point-like $\gamma NN$ coupling $F_1^p=1$ was used. 
In case of pion electroproduction one has $Q^2>0$ and, in principle, $F_1^p=F_1^p(Q^2)$, with $F_1^p(Q^2)$ the Dirac proton form factor which, at low $Q^2$, is well described by a dipole shape $F_1^p(Q^2)\approx\left(1+Q^2/M_V^2\right)^{-2}$.
However, Kaskulov and Mosel argued in Ref.~\cite{Kaskulov10a} that since the intermediate proton might be highly off-shell, it is too naive to use the proton Dirac form factor in the $\gamma NN$ vertex. 
They proposed a transition (off-shell) proton form factor $F_1^p(Q^2,s)$ that absorbs all effects from the virtual proton which oscillates into higher mass and spin resonances. 
This introduces extra strength in the hard sector (large $Q^2$) of the scattering cross section, where the effect of nucleon resonances is expected to be important, and improves the agreement with the electroproduction data in the transverse cross section.
Later on, this idea was employed by Vrancx and Ryckebusch~\cite{Vrancx14a} who proposed an alternative and more intuitive description of $F_1^p(Q^2,s)$. In contrast to the original form factor introduced by Kaskulov and Mosel, the form factor of Ref.~\cite{Vrancx14a} respects the on-shell limit: $F_1^p(Q^2,s)\xrightarrow{s\rightarrow M^2} F_1^p(Q^2)$. 
It is given by
\ba
  F_1^p(Q^2,s) = \left(1 + \frac{Q^2}{\Lambda_{\gamma pp^*}(s)^2}\right)^{-2}\,,\label{eq:F1pregge}
\ea
with
\begin{align}
  \Lambda_{\gamma pp^*}(s) = \Lambda_{\gamma pp} + (\Lambda_\infty-\Lambda_{\gamma pp})\left(1-\frac{M^2}{s}\right).
\end{align}
In the case of the $u$-channel contribution, Eq.~\ref{eq:F1pregge} holds under replacement of $s$ by $u$ with
\ba
  \Lambda_{\gamma pp^*}(u) &=& \Lambda_{\gamma pp} + (\Lambda_\infty-\Lambda_{\gamma pp})\non\\ &\times& \left(1-\frac{M^2}{2M^2 - u}\right).
\ea
$\Lambda_{\gamma pp}\equiv M_V=0.84$ GeV such that $\Lambda_{\gamma pp^*}(M)=\Lambda_{\gamma pp}$ and the on-shell Dirac form factor is recovered. $\Lambda_\infty$ is a free parameter of the model which was fitted to experimental data obtaining $\Lambda_\infty=2.194$ GeV~\cite{Vrancx14a}. 
It was shown by Vrancx and Ryckebusch that this form factor provides predictions quantitatively similar to those obtained by Kaskulov and Mosel.

In case of on-shell initial and final nucleons and $F_2^{p,n}=0$, the gauge-invariant current containing the pion-exchanged $t$-channel diagram and the $s$($u$)-channel amplitude with {\it pseudoscalar} coupling (Eqs.~\ref{eq:JpsNP} and \ref{eq:JpsCNP}), provides exactly the same gauge-invariant current as the ChPT-background model presented in Sec.~\ref{Low-energy-model}, where the EM current is described by the pion-exchanged $t$-channel diagram, the $s$($u$)-channel amplitude with {\it pseudovector} coupling and the $CT$ amplitude (Eq.~\ref{eq:O-ChPT}). 
Since the pion-exchanged $t$-channel diagram is exactly the same in both approaches, one has (for $F_2^{p,n}=0$):
\begin{flalign}
   &\ubarN \left( {\cal O}^{ps}_{NPv} + {\cal O}^{ps}_{CNPv} \right)\uu &&\label{eq:igualdad1}\\ 
  &= \ubarN \left( {\cal O}_{NPv} + {\cal O}_{CNPv} + {\cal O}_{CTv} \right)\uu.&&\non
\end{flalign}
A key ingredient for Eq.~\ref{eq:igualdad1} to be fulfilled is that, for the $p(e,e'\pi^+)n$ reaction, the same proton form factor has to be used in the $CTv$ as in the $NPv$. Analogously, for the $n(e,e'\pi^-)p$ reaction the same proton form factor has to be used in the $CTv$ as in the $CNPv$
\footnote{Note that this is not in contradiction with the fact that the $CTv$ is an isovector operator since $F_1^V = F_1^p$ in the electric model.}.

We shall reggeize the ChPT-background terms by exploiting the ideas described above. 
Thus, by construction, the predictions of the reggeized ChPT-electric background model ({\it ReChi model} in what follows) must be similar to those in Ref.~\cite{Vrancx14a}, which we use as a benchmark.

In what follows, we summarize the ingredients of the ReChi model. 
The vector-current operator is defined by
\ba
  {\cal O}_{ReChi,V}^\mu = {\cal O}_{ChPT,V}^\mu\, {\cal P}_\pi(t,s) (t-m_\pi^2)\,.\label{eq:OReChiV}
\ea
For pion electroproduction, the non-resonant background contributions are simply  
${\cal O}_{ChPT,V}^\mu = {\cal O}_{PF}^\mu + {\cal O}_{NPv}^\mu + {\cal O}_{CNPv}^\mu + {\cal O}_{CTv}^\mu$,
the different terms are the ``electric'' versions ($F_1^n=F_2^{p,n}=0$ and $F_1^p$ given by Eq.~\ref{eq:F1pregge}) of the background contributions described in Eqs.~\ref{eq:ONP}, \ref{eq:OCNP} and \ref{eq:OCTV}.
Also, in the $PF$ term ${\cal O}_{PF}^\mu$, we include the pion transition form factor~\cite{Vrancx14a}
\ba
  F_{\gamma\pi\pi}(Q^2) = \left(1+Q^2/\Lambda_{\gamma\pi\pi}^2\right)^{-1}\,,\label{eq:Ffpipi}
\ea
with pion cut-off parameter $\Lambda_{\gamma\pi\pi}=0.655$ GeV.
${\cal P}_\pi(t,s)$ is the strongly degenerate $\pi(140)/b_1(1235)$-Regge propagator~\footnote{The Regge-propagator in Eq.~\ref{eq:PpiRegge} is obtained in the Regge limit, i.e. high $s$ and small (negative) $t$ values. Therefore, its applicability beyond this region is questionable. 
A detailed derivation can be found in Ref.~\cite{Collins77}.}~\cite{Guidal97,Kaskulov10a,Vrancx14a}
\begin{align}
  {\cal P}_\pi(t,s) = -\alpha_\pi' \varphi_\pi(t)\Gamma[-\alpha_\pi(t)](\alpha_\pi's)^{\alpha_\pi(t)}\,,\label{eq:PpiRegge}
\end{align}
with the Regge trajectory $\alpha_\pi(t)=\alpha'_\pi(t-m_\pi^2)$ and $\alpha_\pi'=0.74$ GeV$^{-2}$. 
The trajectory can be extracted from the pion spectrum. For clarity, one can write $\Gamma[-\alpha_\pi(t)] = -\pi/\{\sin[\pi \alpha_\pi(t)] \Gamma[\alpha_\pi(t)+1]\}$, which contains the pole-generating factor $\sin[\pi \alpha_\pi(t)]$. The $\Gamma[\alpha_\pi(t)+1]$ removes the unphysical contribution of negative-integer spin exchanges. It is interesting to show that the Regge propagator reduces to the pion propagator near the pion pole
\ba
\frac{\pi \alpha_\pi'}{\sin[\pi \alpha_\pi(t)]} \underset{t\to m_\pi^2}{\to} \frac{1}{t-m_\pi^2} \,.\label{eq:PpiReggelowt}
\ea
In Refs.~\cite{Kaskulov10a,Vrancx14a} the Regge phases 
$\varphi_\pi(t) = \exp[-i\pi\alpha_\pi(t)]$ for $p(e,e'\pi^+)n$ 
and $\varphi_\pi(t) = 1$ for $n(e,e'\pi^-)p$ were employed. The choice of phase is related to the relative sign of the degenerate Regge contributions.
Since $\alpha_\pi$ is the only trajectory that we include in our model, the phase is irrelevant.
For that reason, we fix $\varphi_\pi(t) = 1$ for both reaction channels.

\begin{figure*}[htbp]
  \centering  
      \includegraphics[width=.35\textwidth,angle=0]{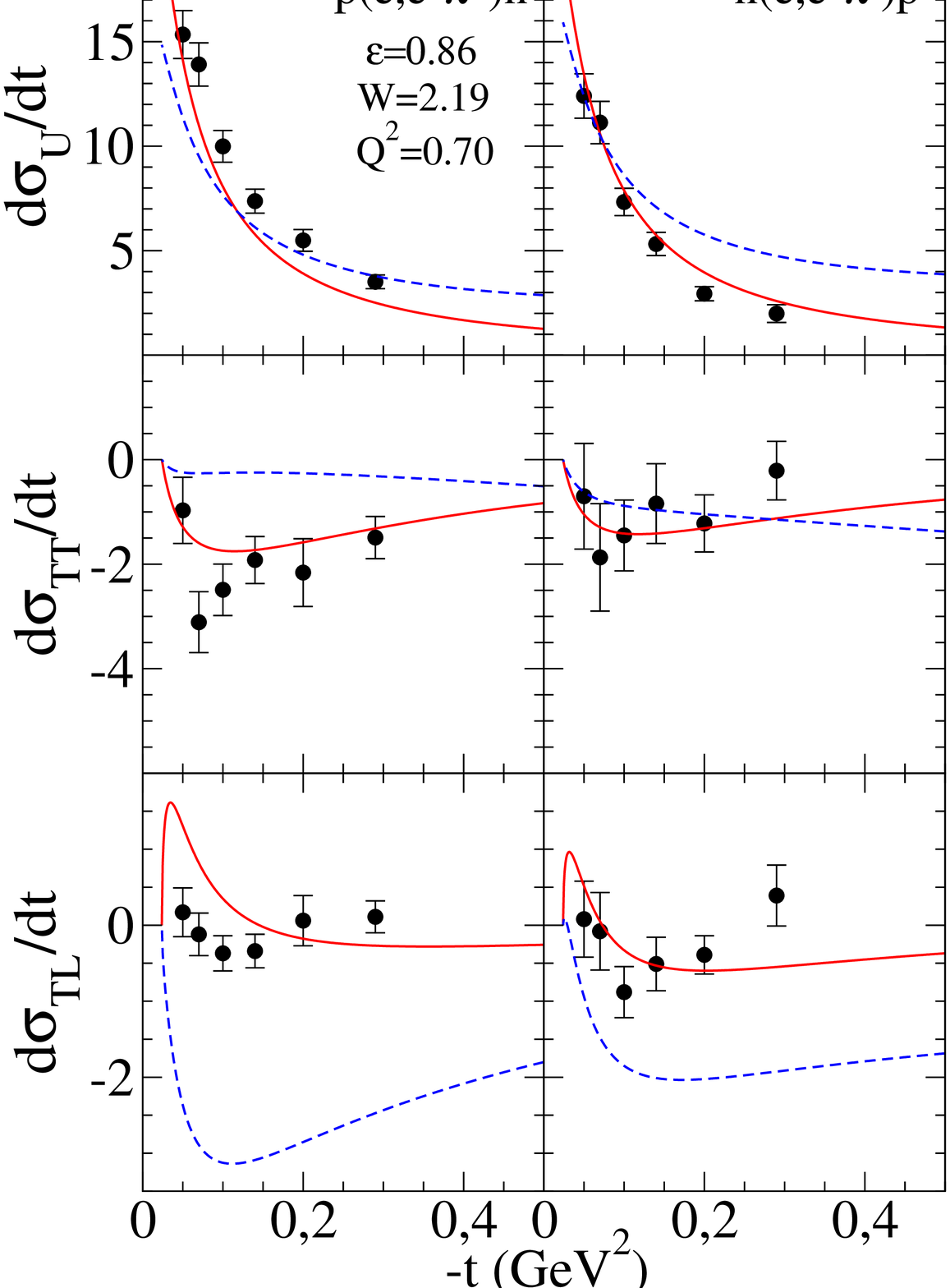}
      \includegraphics[width=.35\textwidth,angle=0]{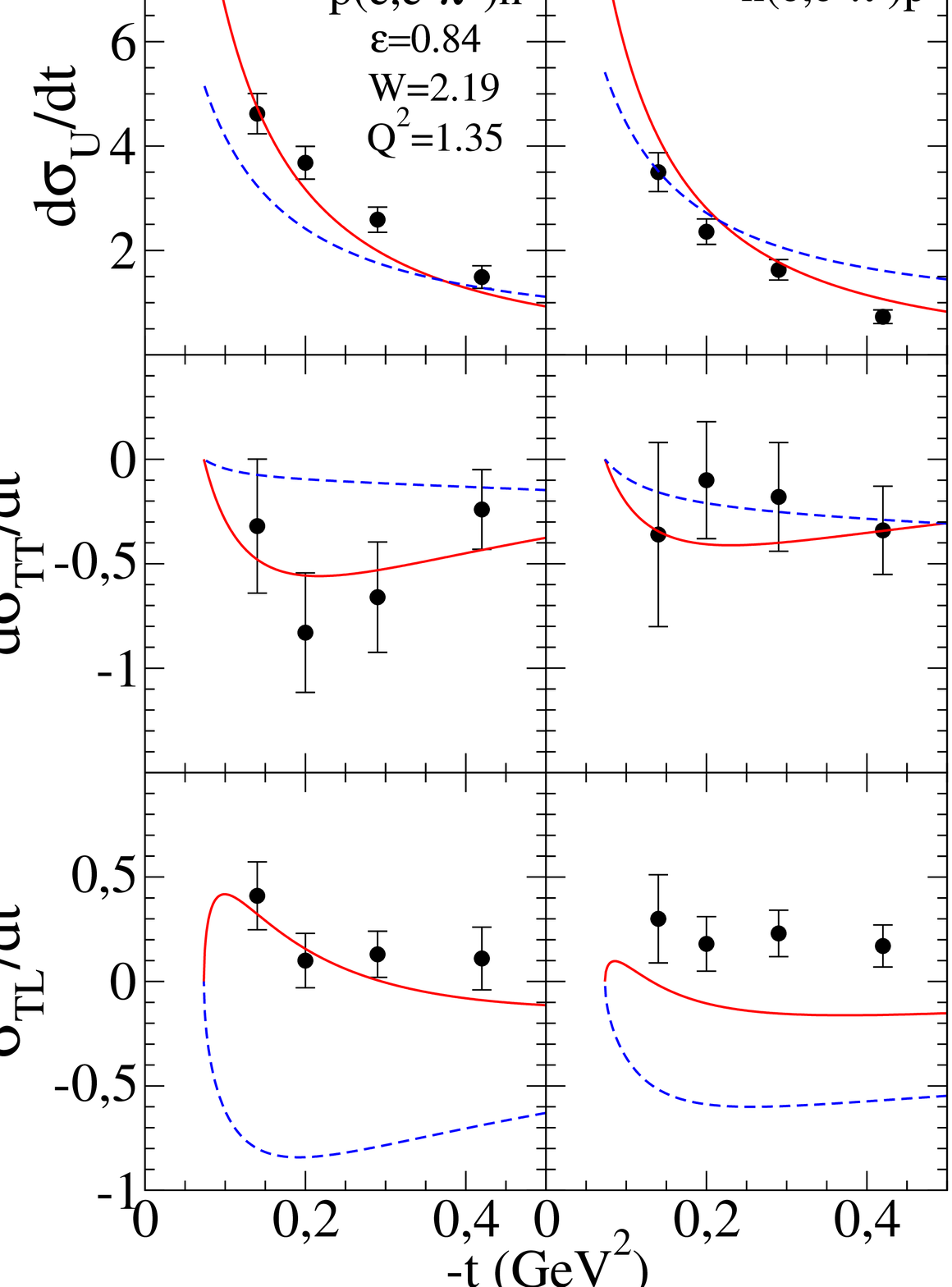}
  \caption{(Color online) Left panels: $U$, $TT$ and $TL$ contributions to the cross section in units of $\mu b/GeV^2$ as a function of $-t$ for fixed $(\epsilon,W,Q^2)$ values. 
  The left and right column correspond to the reactions $p(e,e'\pi^+)n$ and $n(e,e'\pi^-)p$, respectively. 
  The solid-red line is the result of the ReChi model while the dashed-blue line corresponds to the ChPT model.
  The data is from Ref.~\cite{DESY79}.
  Right panels: As before but for different $(\epsilon,W,Q^2)$ values.}
  \label{fig:lowt}
\end{figure*}
\begin{figure*}[htbp]
  \centering  
      \includegraphics[width=.7\textwidth,angle=0]{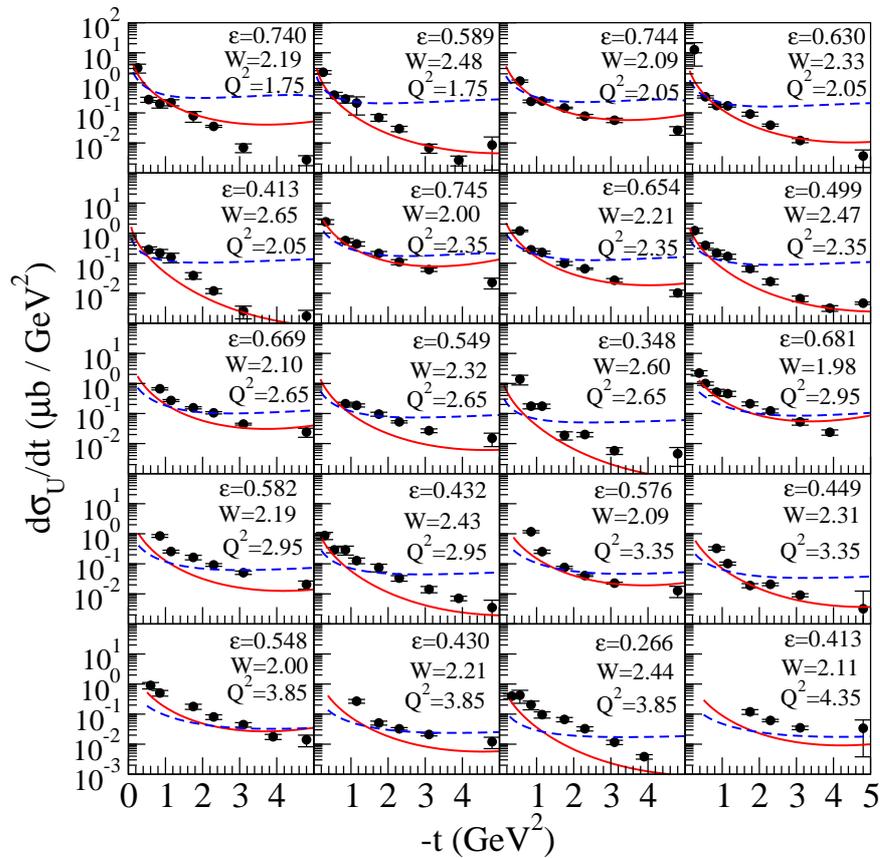}
  \caption{(Color online) $U$ cross section for different $(\epsilon,W,Q^2)$ values as a function of $-t$. 
  The data is from~\cite{CLAS13}. 
  Solid-red and dashed-blue line correspond to the ReChi model and the ChP model, respectively.}
  \label{fig:hight}
\end{figure*}

In Refs.~\cite{Kaskulov10a} and \cite{Vrancx14a}, an ``antishrinkage'' effect which takes into account the decrease of the slope of partonic contributions for increasing values of $Q^2$ was introduced by modifying the slope of the pion trajectory. 
This correction improves the agreement with the transverse data $d\sigma_T/dt$ in the region $-t<0.5$ GeV$^2$ (see Ref.~\cite{Kaskulov10a}). 
On the other hand, it was shown in Ref.~\cite{Vrancx14a} that this model clearly overshoots the experimental data in the region $1<-t<5$ GeV$^2$. 
They found that the predictions can be brought closer to data by fitting a $t$-dependent strong coupling constant $g_{\pi NN}(t)$. 
The use of that $t$-dependent coupling constant, however, notably deteriorates the agreement with data in the low $-t$ region~\cite{Vrancx14a}, where the cross sections are much larger. 
We checked that if the antishrinkage effect is not included, the model presents an acceptable agreement with data in the entire studied region $0<-t<4$ GeV$^2$ without the need of the $t$-dependent coupling $g_{\pi NN}(t)$ (see Figs.~\ref{fig:lowt}-\ref{fig:hight}). Thus, in order to keep the model as basic as possible, we did not include any of these corrections.

In Fig.~\ref{fig:lowt}, the predictions of the ReChi model (solid lines) for the unseparated $d\sigma_U/dt=d\sigma_T/dt+\epsilon d\sigma_L/dt$ and the interference cross sections $d\sigma_{TT}/dt$ and $d\sigma_{TL}/dt$ are compared with the exclusive $p(e,e'\pi^+)n$ and $n(e,e'\pi^-)p$ data in the region $-t<0.5$ GeV$^2$, for two different $Q^2$ values 
\footnote{We have used the same convention as in Refs.~\cite{Kaskulov10a,Vrancx14a} for the definition of $d\sigma_x/dt$ ($x=L,T,TT,TL$) and for the ratio of longitudinal to transverse polarization of the virtual photon $\epsilon$. We refer the reader to Ref.~\cite{Kaskulov10a} for further details.}.
We observe a good agreement with the unseparated $d\sigma_U/dt$ data and a reasonable prediction of the interference $d\sigma_{TT}/dt$ and $d\sigma_{TL}/dt$ cross sections. Actually, our predictions are similar to those of Refs.~\cite{Kaskulov10a,Vrancx14a}, which was expected since they share the main ingredients.
In Fig.~\ref{fig:hight}, we show the cross section $d\sigma_U/dt$ for the process $p(e,e'\pi^+)n$ in the region $0<-t<5$ GeV$^2$, for different fixed values of $Q^2$ and $W$. 
The ReChi model reproduces well the general behavior of the data in the region $-t<4$, where the condition $-t/s<1$ is satisfied. 
The agreement is better in the panels corresponding to lower $Q^2$. For increasing $Q^2$, it seems that the ReChi model systematically underestimates the data. 
As in Fig.~\ref{fig:lowt}, our results are in good agreement with the ones in Ref.~\cite{Vrancx14a}. 
The dashed-blue lines in Figs.~\ref{fig:lowt} and \ref{fig:hight} are the results from the ChPT-background model of Section~\ref{Low-energy-model}. 
This model is practically $t$ independent for $-t>1$ GeV$^2$ and does not reproduce the behavior of the data. For increasing $-t$-values, it clearly overestimates the data, in some cases by several orders of magnitude.

In summary, the predictions of the ReChi model for charged-pion electroproduction are, by construction, quantitatively similar to those from the model of Ref.~\cite{Vrancx14a}, which was used as a benchmark.
Actually, the only differences between the two models are: 
i) we only include the dominant $\pi(140)/b_1(1235)$-Regge trajectory, while in Ref.~\cite{Vrancx14a} (as well as, in Ref.~\cite{Kaskulov10a}) the vector $\rho(770)/a_2(1320)$ and axial-vector $a_1(1260)$ trajectories were also considered, 
ii) we do not include the ``antishrinkage'' effect nor the $t$-dependent strong coupling constant $g_{\pi NN}(t)$.

\begin{figure*}[htbp]
  \centering  
      \includegraphics[width=.45\textwidth,angle=270]{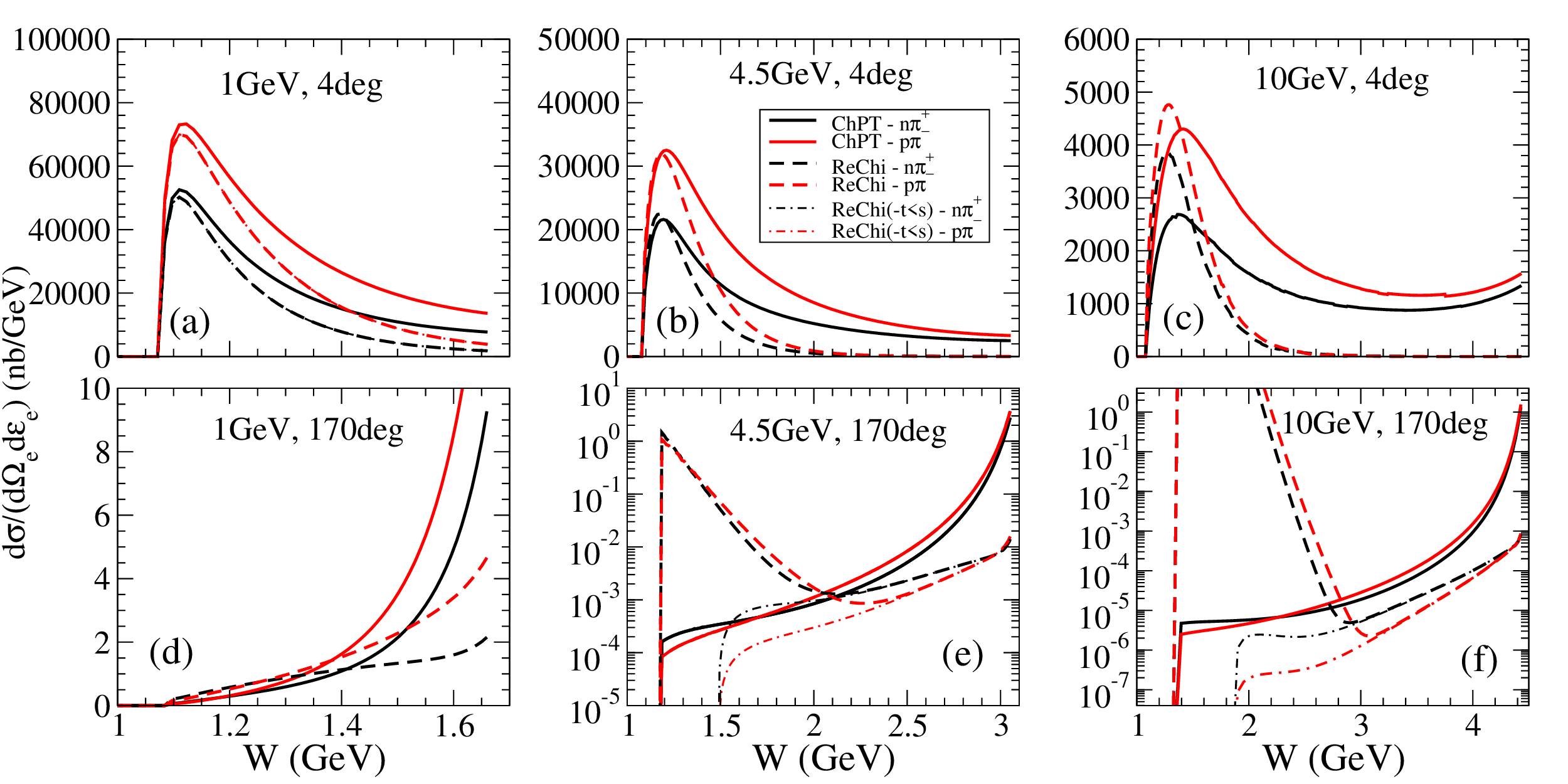}
  \caption{(Color online) Double differential cross section $d\sigma/(d\Omega_e d\varepsilon_e)$ as a function of the invariant mass $W$ for charged-pion electroproduction. 
  The ChPT model (solid lines) is compared with the ReChi model (dashed lines) at six kinematics (see text).
  Black lines correspond to the reaction $p(e,e'\pi^+)n$ and red lines to $n(e,e'\pi^-)p$.
  Dashed-dotted lines in panels (e) and (f) correspond to the predictions of the ReChi model when the kinematic cut $-t/s<1$ is applied.}
  \label{fig:bgs-electrons}
\end{figure*}

In Fig.~\ref{fig:bgs-electrons}, we compare the ReChi model (dashed lines) with the ChPT model (solid lines). 
We show the double differential cross section $d\sigma/(d\Omega_e d\varepsilon_e)$ for the set of kinematics studied in Sec.~\ref{Kinematic}.
Note that, for a given $W$ value, an integral over $t$ is performed and, therefore, all $t$ values allowed by kinematics contribute (see Fig.~\ref{fig:tregion}). 
In what follows, we qualitatively analyze the behavior of the ReChi and ChPT models in the possible kinematic scenarios.

{\em Low-energy region:} $W\lesssim1.4$ GeV. This is the region where the ChPT model is reliable. 
Our Regge-based model should not be applied here for the reasons explained below.
At small $-t$, the Regge propagator tends to the pion propagator (Eq.~\ref{eq:PpiReggelowt}), therefore, the predictions of the reggeized and non-reggeized models tend to be similar. 
In spite of that, the ReChi model, by construction, lacks some ingredients that are relevant at low $W$, such as the possibility of coupling to the neutron ($F_1^n=0$) and the contribution from the anomalous tensor coupling ($F_2^{p,n}=0$). 
For that reason, in this regime the ChPT model is preferable to the ReChi model.
In Fig.~\ref{fig:bgs-electrons}, the results of panels (a), (b), (c) and (d) in the region $W<1.4$ GeV fit in this situation. Note that only small $-t$ values contribute (see Fig.~\ref{fig:tregion}). 
At large $-t$, one has $-t/s>1$ and we are far from the Regge limit. The ReChi model is not valid in this situation.
This is the case in the low-$W$ region of panel (e) and (f), where the ReChi model provides non-sense cross sections. 

{\em High-energy region:} $W\gtrsim2$ GeV.
Regge-based models are a good alternative in this region, where low-energy models fail.
At small $-t$, we enter in the pure Regge limit ($-t/s<<1$). This is the natural domain of Regge-based models.
For increasing $-t$ values, however, one may enter the region $-t/s\lesssim1$.  
The predictions of the ReChi model are less reliable than in the previous situation, but still preferable to the ChPT model.
One expects that the ReChi model slightly underestimates the $t$-integrated cross sections due to the lack of the $u$-channel baryon-exchange contribution at backward $\theta_\pi^*$ scattering. 
In panels (e) and (f) of Fig.~\ref{fig:bgs-electrons}, we show the predictions of the ReChi model when the kinematic condition $-t/s<1$ is applied (dashed-dotted lines)~\footnote{Note that the kinematic cut $-t/s<1$ does not affect the results in panels (a), (b), (c) and (d) in Fig.~\ref{fig:bgs-electrons} (see Fig.~\ref{fig:tregion}).}.
As expected, the unphysically high responses observed in the low-$W$ region of panels (e) and (f) disappear. Also, the predictions with and without the kinematic cut overlap from a certain $W$ value, which could be expected since low-$t$ contributions strongly dominate the cross sections. 
Thus, we will use the condition $-t/s<1$ as the limit of applicability for the ReChi model.
We want to stress that, provided the condition $-t/s\lesssim1$ is fulfilled, the ReChi model works reasonably well even at relatively low $W$ ($W\approx2$ GeV), while the ChPT model clamorously fails (Figs.~\ref{fig:lowt} and \ref{fig:hight}).

{\em Transition region:} $1.4\lesssim W \lesssim 2$ GeV. The pathologies of the low-energy models become manifest in this region. Also, this is not the natural domain of Regge-based models. 
In a phenomenological sense, one may consider that, in this transition region, more realistic results could arise from a compromise between the low-energy and the Regge-based predictions.

\subsection{Neutrinoproduction of pions}\label{regge-neutrinos}

The ReChi model presented above for EM interaction is extended here to CC and WNC neutrino-induced pion production.
In this case, the non-resonant current operator contains a vector (V) and an axial (A) contribution:
\ba
{\cal O}_{ChPT,V}^\mu = {\cal O}_{NPv}^\mu + {\cal O}_{CNPv}^\mu + {\cal O}_{PF}^\mu + {\cal O}_{CTv}^\mu, &\label{eq:OreggeV}\non\\
{\cal O}_{ChPT,A}^\mu = {\cal O}_{NPa}^\mu + {\cal O}_{CNPa}^\mu + {\cal O}_{CTa}^\mu + {\cal O}_{PP}^\mu. &\label{eq:OreggeA}\non\\
\ea

\subsubsection{Reggeizing the vector current}

Based on CVC and by isospin rotation, the vector current for neutrinoproduction can be determined from the electroproduction current. 

In the case of the WNC interaction, the nucleon form factors are given in terms of the EM ones by Eq.~\ref{eq:WNC-ff}. 
Hence, under the assumption (electric model) $F_1^n=F_2^{p,n}=0$, one obtains:
\ba
  \widetilde{F}_1^p[Q^2,s(u)] &=& \frac{1}{2}(1-4\sin^2\theta_W) F_1^p[Q^2,s(u)]\,,\non\\
  \widetilde{F}_1^n[Q^2,s(u)] &=& -\frac{1}{2}F_1^p[Q^2,s(u)]\,.
\ea
This means that, contrary to the situation in charged-pion electroproduction, the Z boson also couples to neutrons and, therefore, both $NPv$ and $CNPv$ contribute to the reactions $p(\nu,\nu'\pi^+)n$ and $n(\nu,\nu'\pi^-)p$.
Taking that into account, it is easy to define the form factor that enters in the $CTv$ amplitude. 
For $p(\nu,\nu'\pi^+)n$ one has
\begin{flalign}
  \widetilde{F}^{n\pi^+}_{CT}(Q^2,s,u) = \frac{1}{2}[\widetilde{F}_1^p(Q^2,s) - \widetilde{F}_1^n(Q^2,u)],
\end{flalign}
while for $n(\nu,\nu'\pi^-)p$ one obtains
\begin{flalign}
  \widetilde{F}^{p\pi^-}_{CT}(Q^2,u,s) = \frac{1}{2}[\widetilde{F}_1^p(Q^2,u) - \widetilde{F}_1^n(Q^2,s)].
\end{flalign}

In the case of the CC interaction, we need the isovector form factors for the $CTv$, $NPv$ and $CNPv$ amplitudes. In the electric model these are: $F_1^V=F_1^p$, $F_2^V=0$. 
Additionally, for CC neutral-pion production, both $NPv$ and $CNPv$ amplitudes contribute (see table~\ref{Iso-coef-CC}). In this case,
Eq.~\ref{eq:igualdad1} implies that the form factor that enters in the $CTv$ amplitude has to be:
\begin{flalign}
  F^{\pi^0}_{CT}(Q^2,s,u) = \frac{1}{2}[F_1^p(Q^2,s) + F_1^p(Q^2,u)].
\end{flalign}

Finally, the ReChi vector-current operator for neutrino-induced SPP takes the form of Eq.~\ref{eq:OReChiV}.

\subsubsection{Reggeizing the axial current}

The presence of the $PF$ diagram in the vector current enabled us to apply the reggeizing procedure by identifying the pion exchange as the main Regge trajectory.
However, what is the analogous $t$-channel meson-exchange diagram in the axial current? 

\begin{figure}[htbp]
  \centering  
      \includegraphics[width=.5\textwidth,angle=0]{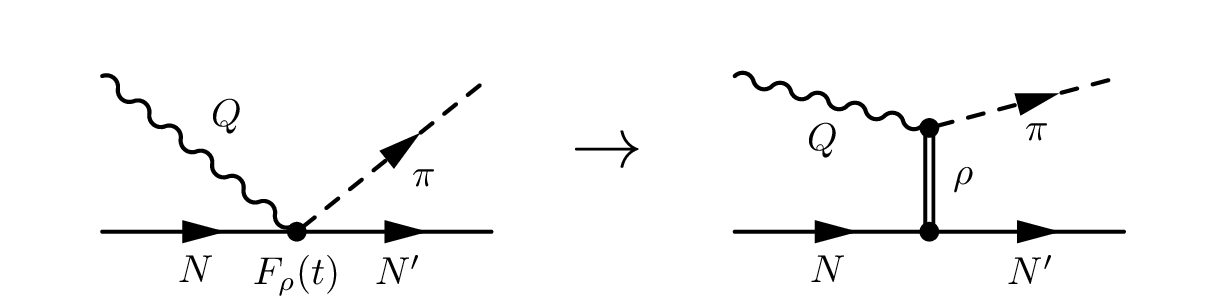}\\
      \includegraphics[width=.5\textwidth,angle=0]{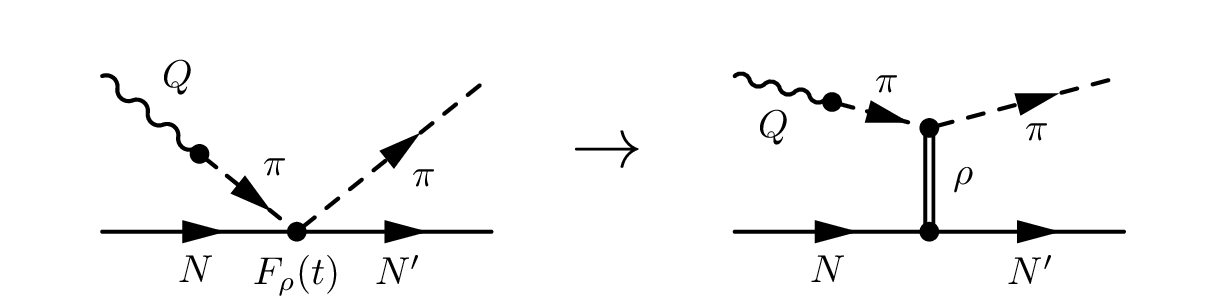}
  \caption{In the left side we present the $CTa$ and $PP$ diagrams used in the low-energy model of Sec.~\ref{Low-energy-model}. 
  In the high-energy model we reinterpret these diagrams as $\rho$-exchange diagrams (right side): $CT\rho$ and $PP\rho$.
  In this figure, $Q$ represents the axial-vector part of the weak  ($W^\pm$ or $Z$) boson.}
  \label{fig:rho-exchanged}
\end{figure}

In the $\pi\pi NN$ and $Q\pi NN$ vertices which are present in the $PP$ and $CTa$ diagrams, the form factor $F_\rho(t)$ was introduced
to, phenomenologically, account for the $\rho$-dominance of the $\pi\pi NN$ vertex (see Sec.~\ref{Low-energy-model}).
Actually, the $CTa$ and the $PP$ diagrams with this form factor can be interpreted as effective $\rho$-exchange diagrams. 
This is illustrated in Fig.~\ref{fig:rho-exchanged}~\footnote{The $\rho$-exchange contribution to the axial current has been previously considered in the low-energy models of Refs.~\cite{Towner92,Sato03}, and in the high-energy models of Refs.~\cite{Gershtein80,Rein86,Allen86}.}.
Thus, we identify the $\rho$ exchange as the main Regge trajectory in the axial current, this will allow us to reggeize the axial current by using the $\rho$-exchanged Regge propagator.

The current operator for the $CT\rho$ in Fig.~\ref{fig:rho-exchanged} is
\ba
 {\cal O}^\mu_{CT\rho} &=& i{\cal I}\, \frac{m_\rho^2}{m_\rho^2-t} F_{A\rho\pi}(Q^2) \frac{1}{\sqrt2f_\pi}\non\\
 &\times&\left(\gamma^\mu + i\frac{\kappa_\rho}{2M}\sigma^{\mu\nu}K_{t,\nu}\right) \,. \label{eq:OrhoCTA}
\ea
For the coupling constants, we have assumed~\cite{Sato03} 
\ba
  g_{\rho NN}g_{A\rho\pi} = m_\rho^2/f_\pi\,,
\ea
$g_{\rho NN}$ and $g_{A\rho\pi}$ being the coupling constants of the strong and weak vertices, respectively.
We also introduced a transition form factor in the weak vertex, $F_{A\rho\pi}(Q^2)$, in analogy with what was done in the $PF$ diagram (see Eq.~\ref{eq:Ffpipi}).  
Within a meson-dominance framework, the axial-vector part of the $W^\pm$ (or $Z$) boson transforms into an $a_1(1260)$ axial-vector meson~\cite{Lichard97}. This suggests the form factor  
\ba
  F_{A\rho\pi}(Q^2) = (1+Q^2/\Lambda_{\rho}^2)^{-1}\,,
\ea
with $\Lambda_{\rho}=m_{a_1(1260)}$.

The $PP$ contribution to the cross section is generally small, because its amplitude is proportional to $Q^\mu$: for WNC interactions it vanishes when it is contracted with the leptonic tensor, while for CC interactions one gets a contribution proportional to the squared mass of the charged lepton. 
Still, the $PP$ amplitude is needed to preserve PCAC, and we keep it in our calculations~\footnote{It is shown in Ref.~\cite{Rein07} that the contribution from the $PP$ term is important for CC pion production in the low $Q^2$ region since it may partially explain the deficit of forward-going muons observed at $Q^2<0.1$ GeV$^2$ in the K2K experiment~\cite{K2K05}.}.
Using PCAC, the $PP\rho$ current operator results 
\ba
 {\cal O}^\mu_{PP\rho} &=& i{\cal I}\, \frac{m_\rho^2}{m_\rho^2-t} F_{A\rho\pi}(Q^2) \frac{-1}{\sqrt2f_\pi} \frac{Q^\mu}{Q^2-m_\pi^2}\non\\
 &\times&\frac{(Q + K_\pi)_\nu}{2} 
  \left( \gamma^\nu + i\frac{\kappa_\rho}{2M}\sigma^{\nu\alpha}K_{t,\alpha} \right).\label{eq:OrhoPP}
\ea

We now follow the same steps as in the case of charged-pion electroproduction and introduce an off-shell axial form factor in the $NPa$ and $CNPa$ amplitudes.
By analogy with the off-shell proton form factor $F_1^p[Q^2,s(u)]$ of Eq.~\ref{eq:F1pregge}, we propose
\ba
  G_A[Q^2,s(u)] = g_A \left(1 + \frac{Q^2}{\Lambda_{A pn^*}[s(u)]^2}\right)^{-2}\label{eq:GAregge}
\ea
with 
\ba
  \Lambda_{A np^*}(s) = \Lambda_{A pn} + (\Lambda^A_\infty-\Lambda_{A pn})\left(1-\frac{M^2}{s}\right),\non\\
\ea
and
\ba
  \Lambda_{A pn^*}(u) &=& \Lambda_{A pn} + (\Lambda^A_\infty-\Lambda_{A pn})\non\\
  &\times&\left(1-\frac{M^2}{2M^2 - u}\right)\,.
\ea 
In order to recover the on-shell axial form factor for $\Lambda_{A np^*}(M)$, we use $\Lambda_{A pn}\equiv M_A=1.05$ GeV.
$\Lambda^A_\infty$ is a free parameter of the model.

While in the vector current, CVC tells us that the $NPv$ and $CNPv$ amplitudes have to be multiplied by the same Regge propagator as the $PF$ diagram, in the axial current there is no such constraint: the $NPa$ and $CNPa$ fulfill PCAC by themselves as well as the combination of $CT\rho$ and $PP\rho$.
Thus, just by analogy with the procedure followed in the vector-current case, we multiply the $NPa$ and $CNPa$ with the same Regge propagator used in the meson-exchange diagrams ($CT\rho$ and $PP\rho$). 
Other options may be explored. 

In summary, the axial-current operator in the ReChi model reads
\ba
  {\cal O}_{ReChi,A}^\mu = {\cal O}_{ChPT,A}^\mu\, {\cal P}_\rho(t,s) (t-m_\rho^2)\,,\label{eq:OReChiA}
\ea
where ${\cal O}_{ChPT,A}^\mu$ is given in Eq.~\ref{eq:OreggeA} and ${\cal P}_\rho(t,s)$ is the strongly degenerate $\rho(770)/a_2(1320)$-Regge propagator~\cite{Guidal97,Kaskulov10a,Vrancx14a}
\begin{align}
  {\cal P}_\rho(t,s) = -\alpha_\rho' \varphi_\rho(t)\Gamma[1-\alpha_\rho(t)](\alpha_\rho's)^{\alpha_\rho(t)-1}.\label{eq:PrhoRegge}
\end{align}
The $\rho$ trajectory is parameterized by $\alpha_\rho(t)=0.53 + \alpha'_\rho t$ with $\alpha_\rho'=0.85$ GeV$^{-2}$. 

Since we are considering a strongly degenerate trajectory, we should choose between a rotating or a constant phase, $\varphi_\rho(t)$.
In case of inclusive cross sections (in which the information about the hadronic system is integrated), it can be shown from general symmetry arguments that the vector/axial (VA) interference contribution to the hadronic tensor is a purely antisymmetric tensor~\cite{Donnelly85}. As a consequence, the VA contribution to the cross section arises from its contraction with the antisymmetric leptonic tensor, $a_{\mu\nu}$ (Eq.~\ref{eq:amunu}). On the other hand, it is easy to show that, at high energies and very forward lepton scattering angles, $a_{\mu\nu}$ vanishes. Therefore, one expects the VA contribution to be considerably smaller than the axial/axial (AA) and vector/vector (VV) ones.
This, together with the fact that $\alpha_\rho$ is the only trajectory included in the axial current of our model, allows us to conclude that a constant or rotating phase will produce basically the same squared amplitudes. We fix $\varphi_\rho(t) = 1$.

The only difference between Eqs.~\ref{eq:OCTA} and \ref{eq:OPP} and Eqs.~\ref{eq:OrhoCTA} and \ref{eq:OrhoPP} is the contribution in the latter of the term proportional to the anomalous tensor coupling constant $\kappa_\rho$, and the transition form factors $F_{A\rho\pi}(Q^2)$. 
Therefore, in this work we will assume $\kappa_\rho=0$~\footnote{One can find in the literature a diversity of values for $\kappa_\rho$, for instance, $\kappa_\rho\approx2$~\cite{Sato96}, $\kappa_\rho=3.71$~\cite{Gari76}, $\kappa_\rho=6.1$~\cite{Machleidt87}.}, such that the background amplitudes of the original low-energy model are recovered in the limits $t\rightarrow m_\rho^2$ and $Q^2\rightarrow0$.
Other approaches may be investigated.

\subsubsection{Results}

\begin{figure*}[htbp]
  \centering  
      \includegraphics[width=.35\textwidth,angle=270]{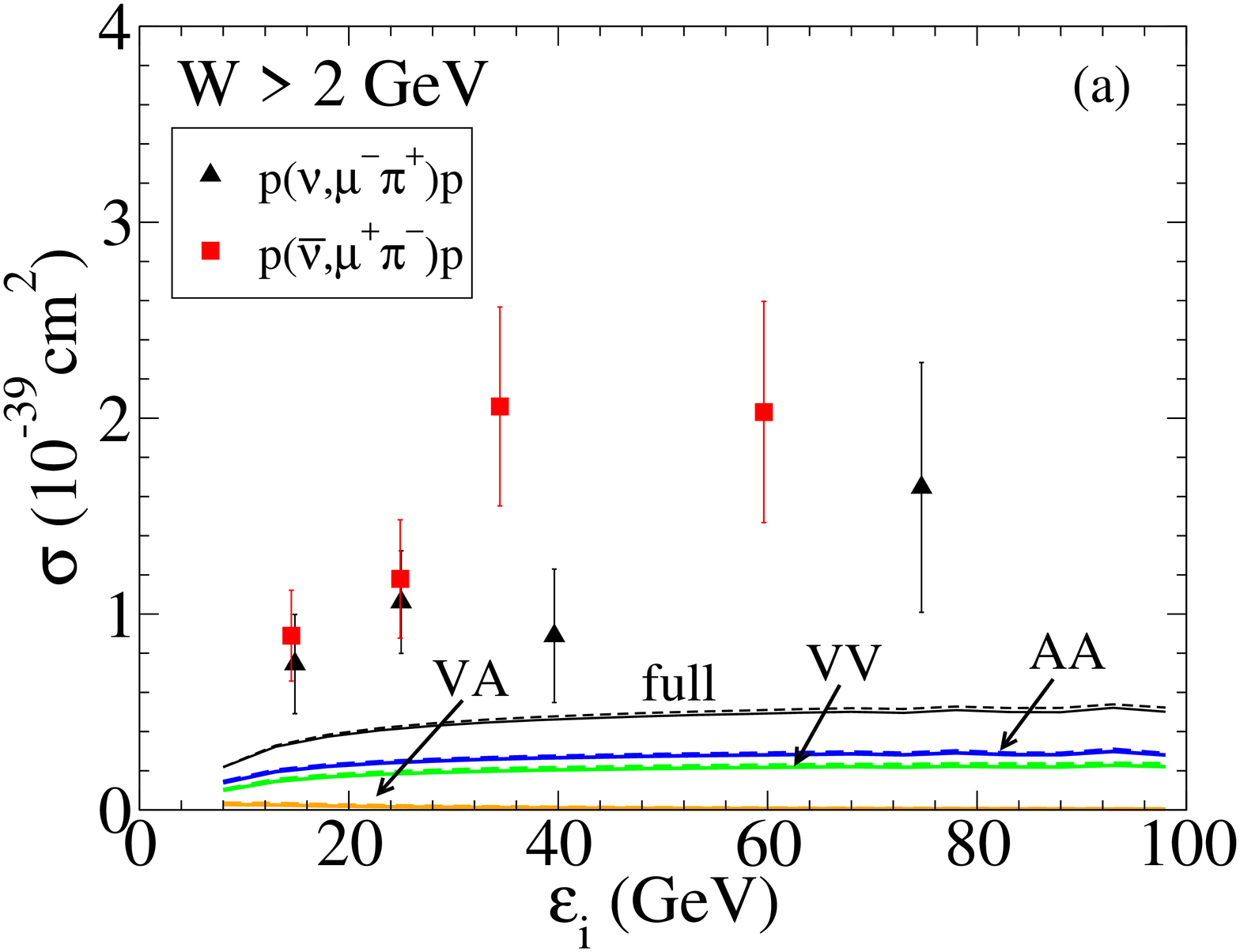}
      \includegraphics[width=.35\textwidth,angle=270]{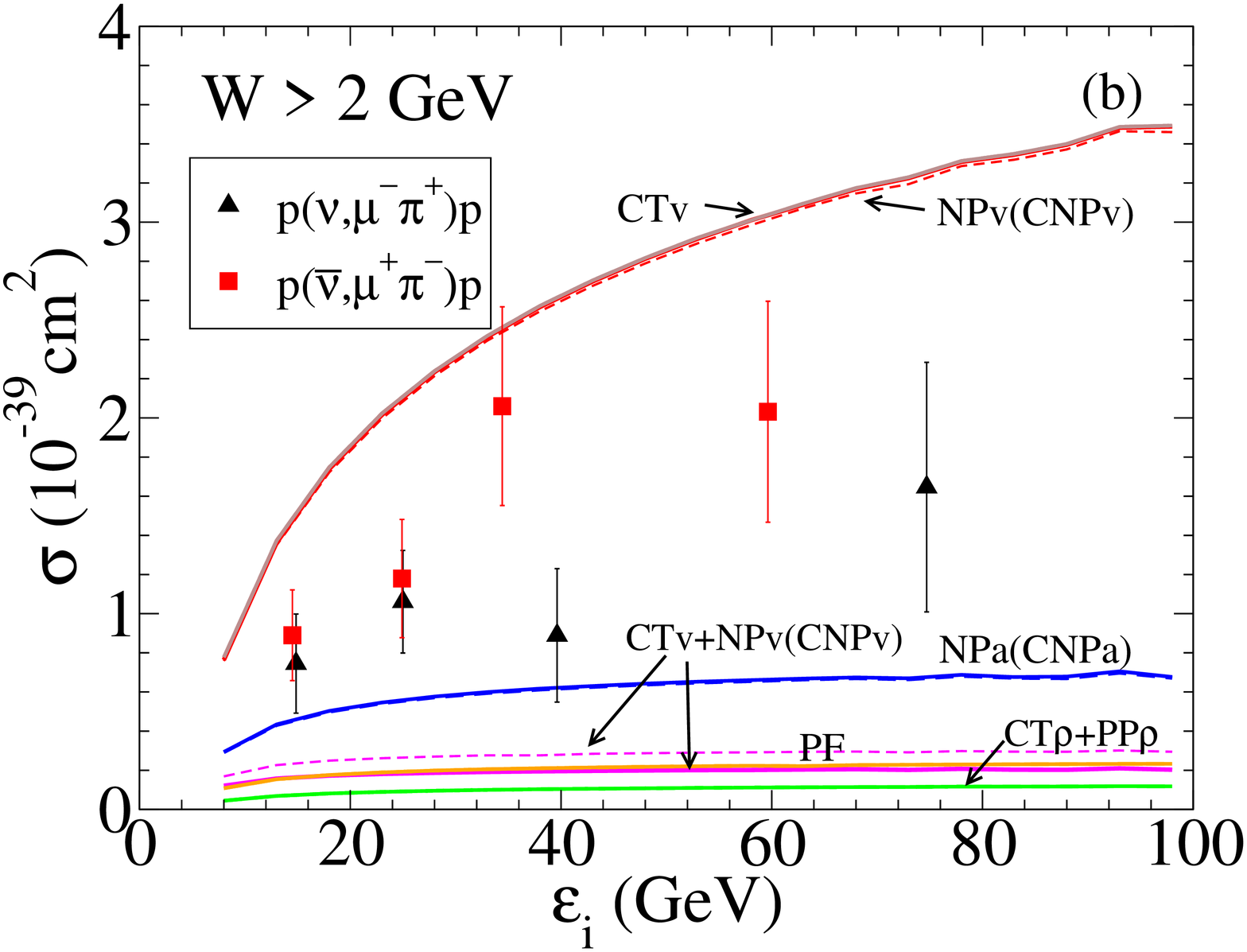}
  \caption{(Color online) Total cross section data for the reactions $\bar\nu p\rightarrow\mu^+\pi^- p$ and $\nu p\rightarrow\mu^-\pi^+ p$ are compared with the predictions from the ReChi model. In both panels (a) and (b), the solid and dashed lines are, respectively, the neutrino and antineutrino predictions (the lines coincide in almost all cases). Data is from Ref.~\cite{Allen86}.}
  \label{fig:CC-high-2194}
\end{figure*}

In Fig.~\ref{fig:CC-high-2194}, we compare the ReChi model with total cross section data~\cite{Allen86} for the CC reactions 
$\bar\nu p\rightarrow\mu^+\pi^- p$ and $\nu p\rightarrow\mu^-\pi^+ p$ in the energy range between 10 and 90 GeV.
The data and the predictions include the kinematic condition $W>2$ GeV, so one expects that only residual effects from the resonance region affect the data. Thus, this is an excellent opportunity to test the ReChi model.
Additionally, we have applied the condition $-t/s<1$ in our model (see Sec.~\ref{regge-electrons}).

\begin{figure*}[htbp]
  \centering  
      \includegraphics[width=.35\textwidth,angle=270]{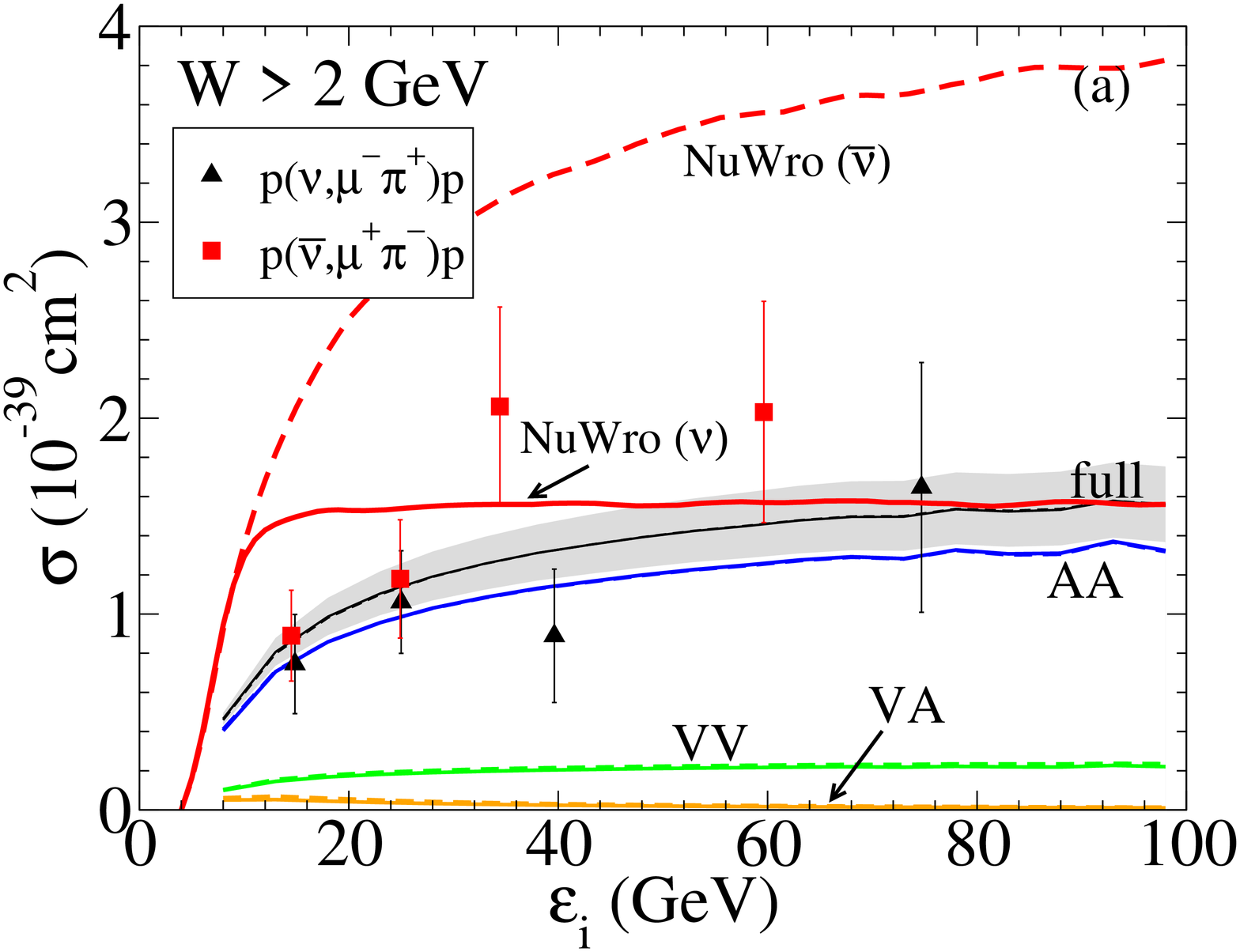}
      \includegraphics[width=.35\textwidth,angle=270]{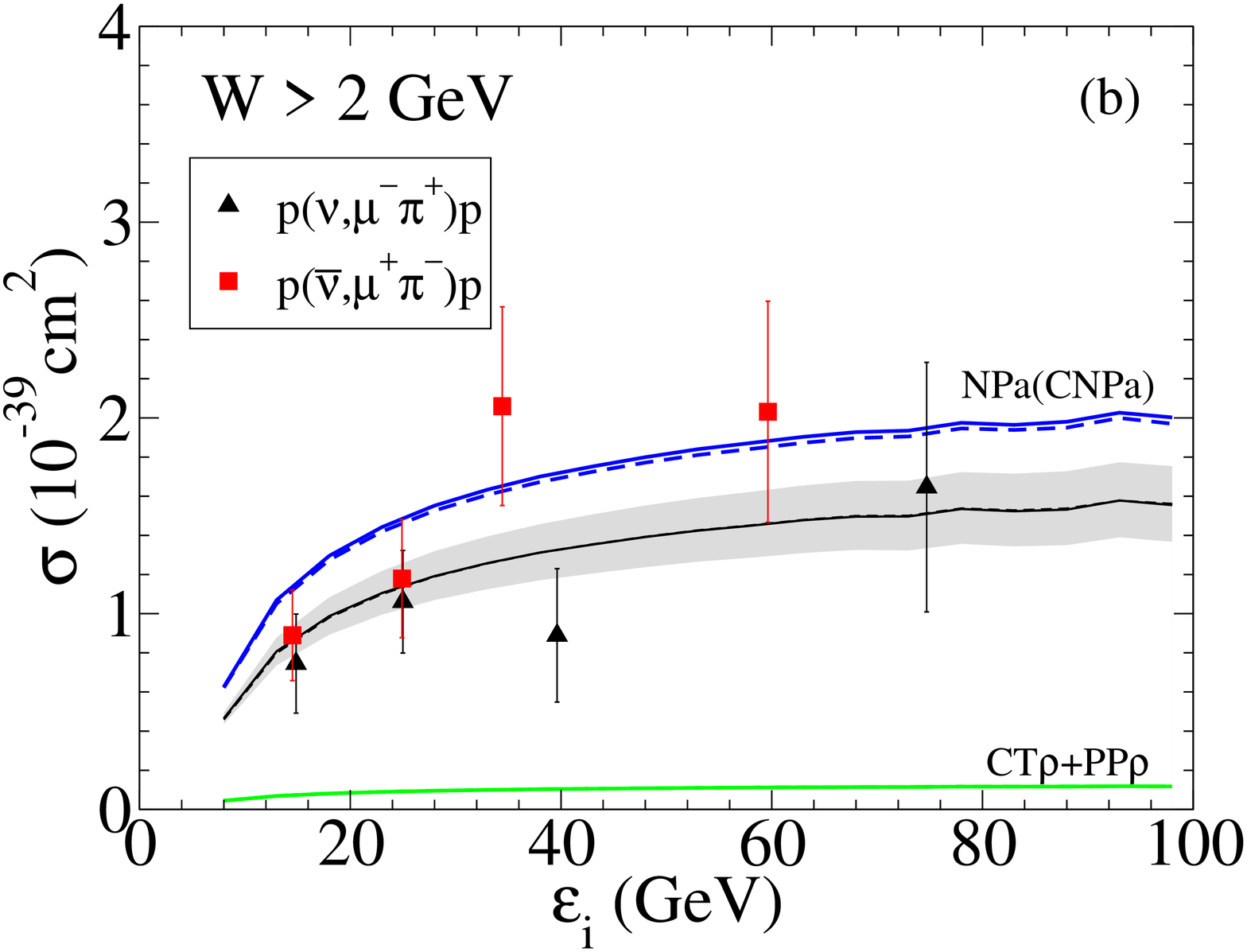}
  \caption{(Color online) As Fig.~\ref{fig:CC-high-2194} but using $\Lambda_\infty^A$ from Eq.~\ref{eq:LinfA} (the gray band represents the 1$\sigma$ region). In panel (a) the NuWro predictions are included (see text for details).}
  \label{fig:CC-high}
\end{figure*}

For the free parameter in the transition axial form factor $\Lambda_\infty^A$, we have used, as reference, the same value as in the analogous parameter of the vector current, i.e. $\Lambda_\infty^A = \Lambda_\infty = 2.194$ GeV. 
In this case, the ReChi model underestimates the data by more than a factor two [``full'' in panel (a)].
In panel (a), we show the VV, AA and VA contributions separately. The VV and AA contributions are similar while, as expected, the VA one is significantly smaller.

In panel (b), we show the contributions to the cross section from the different diagrams of the model: the $NPv$ ($CNPv$), the $CTv$ and the $PF$ amplitudes in the VV sector; the $CT\rho+PP\rho$ and the $NPa$ ($CNPa$) amplitudes in the AA sector. 
It is interesting that, separately, the $NPv$ ($CNPv$) and the $CTv$ clearly overshoot the data, while the combination $CTv+NPv$ ($CNPv$) lays far below data. This is a consequence of the destructive interference between the different contributions. Note that the interferences between all the diagrams are determined by the (low-energy) ChPT model.

In Fig.~\ref{fig:CC-high}, we present the same results as in Fig.~\ref{fig:CC-high-2194} when the parameter
\ba
  \Lambda_\infty^A = \left(7.20 \pm \text{}^{2.09}_{1.32}\right) \text{GeV}\,\label{eq:LinfA}
\ea
is employed. 
This is the results of a $\chi^2$ fit to the 8 experimental data points. The errors in $\Lambda_\infty^A$ define the $1\sigma$ region ($\chi^2<\chi^2_{min}+1$). 
In this case, our result is quantitatively similar to the one in Ref.~\cite{Rein86}.
By increasing $\Lambda_\infty^A$ one obtains a much larger AA contribution [panel (a)], which is a consequence of the augment of the $NPa$ ($CNPa$) term [panel (b)]. This increment of the $NPa$ ($CNPa$) term breaks the equilibrium that exist between the $CT\rho+PP\rho$ and the $NPa$ ($CNPa$), which translates into a net increment of the cross section.

The large value $\Lambda_\infty^A=7.20$ GeV which is needed to reproduce the data, may be a consequence of the lack of other missing ingredients in the model. For instance, the contributions of other meson-exchange trajectories in both the axial and the vector currents may increase the magnitude of the cross section. Note that, as shown in Fig.~\ref{fig:hight}, the ReChi model systematically underestimates the charged-pion electroproduction data for $Q^2\gtrsim2.5$ GeV$^2$. Also, a proper modeling of the backward $\theta_\pi^*$ scattering cross section is missing. Another possibility is, simply, a wrong interpretention of the experimental data, which may contain contributions beyond the purely one-pion production process.

In panel (a) of Fig.~\ref{fig:CC-high}, we included the predictions of the NuWro Monte Carlo event generator~\cite{NuWro-web}. 
In NuWro, in the kinematical regime $W>2$ GeV, the inclusive cross sections are evaluated using the DIS formalism by Bodek-Yang~\cite{Bodek02} and the hadronic final states are obtained using PYTHIA 6 hadronization routines~\cite{Pythia6} (see Ref.~\cite{Sobczyk05,Nowak06} for details). 
The SPP channel is defined by selecting those events with only one pion and one nucleon in the final state.
The NuWro predictions for the channel $\bar\nu p\rightarrow\mu^+\pi^- p$ are approximately a factor 2.3 larger than those for the channel $\nu p\rightarrow\mu^-\pi^+ p$, contrary to the predicitons of the ReChi model that produces almost identical results for the two channels.
As mentioned, the NuWro predictions are based on the DIS formalism, which uses quarks as degrees of freedom. 
Thus, a possible explanation for this factor $\sim$2 arises from the fact that 
a proton consists of two up quarks and one down quark, 
which couple to the antineutrino and the neutrino, respectively:
% that are the needed targets for the antineutrino and the neutrino channels, respectively:
%, therefore,
%the probability for the antineutrino reaction occurs is twice that for the neutrino one:
\ba
  \bar\nu + \overbrace{uud}^{p} \rightarrow \mu^+ + \overbrace{\bar u d}^{\pi^-} + uud\,,\\
  \nu + uud \rightarrow \mu^- + \underbrace{u \bar d}_{\pi^+} + uud\,.  
\ea
% the antineutrino can couple two up quarks of the proton target while the neutrino can couple only one down quark:
The ratio of antineutrino to neutrino data in Fig.~\ref{fig:CC-high} is close to one, which may be interpreted as a sign that DIS is not the dominant reaction mechanism. 
A different problem is that the NuWro predictions overestimate the data, except for the highest energy data point for the neutrino reaction. 

We conclude that further investigations are needed, especially, when more recent and differential cross section data become available.

\begin{figure}[htbp]
  \centering  
      \includegraphics[width=.45\textwidth,angle=0]{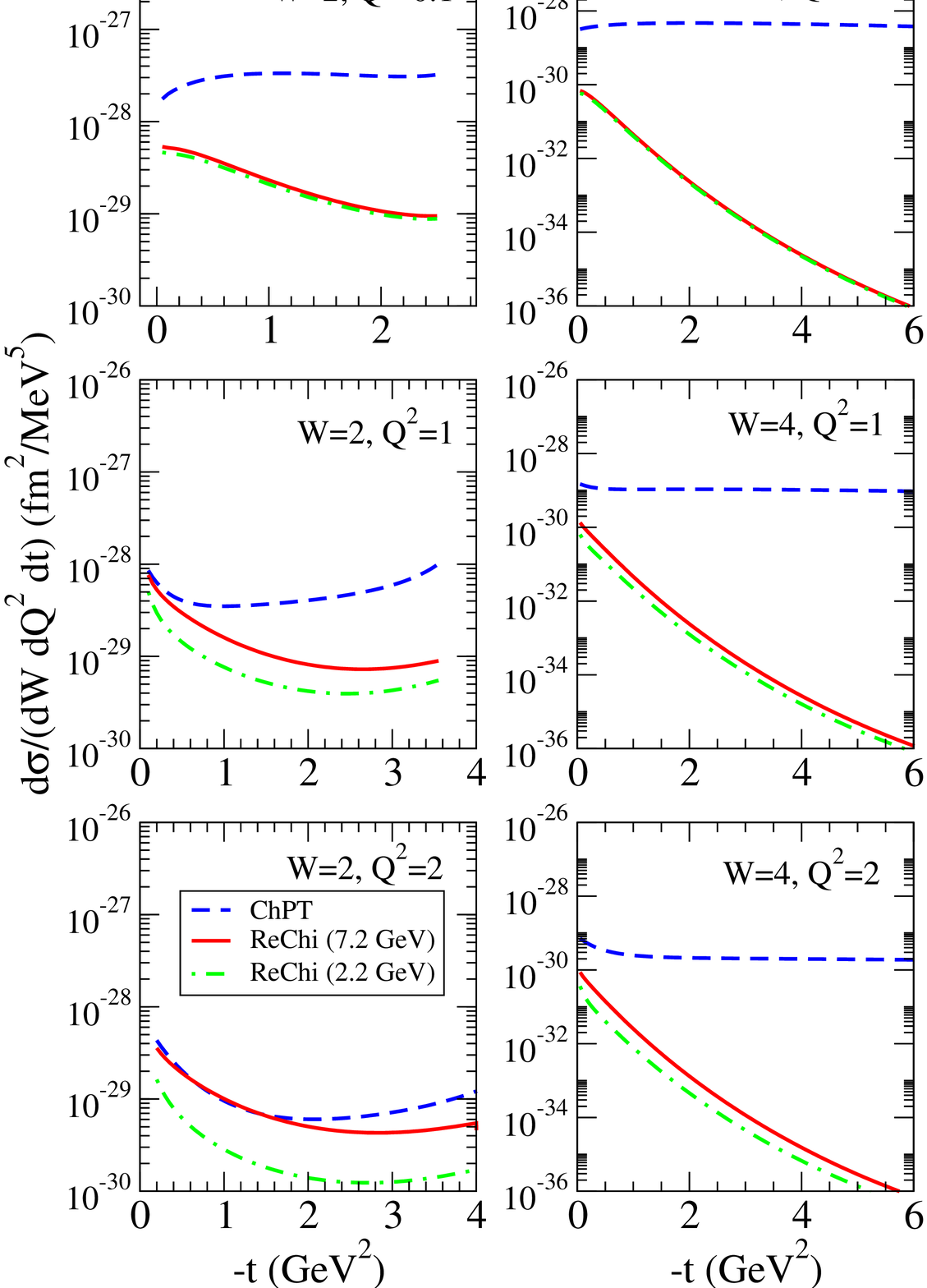}
  \caption{(Color online) The solid-black and solid-thin red lines corresponds to the ReChi model with the values $\Lambda_\infty^A=7.2$ and 2.2 GeV, respectively. The blue-dashed line is the ChPT model.}
  \label{fig:ds-dQ2dWdt}
\end{figure}

In Fig.~\ref{fig:ds-dQ2dWdt}, we further explore the effect of the parameter $\Lambda_\infty^A$ on the cross sections. 
To this end, we show the differential cross section $d\sigma/(dWdQ^2dt)$ as a function of $t$ for some fixed values of $Q^2$ and $W$.
The predictions of the ReChi model were obtained with the values $\Lambda_\infty^A=7.2$ and 2.2 GeV.
The effect of the parameter $\Lambda_\infty^A$ is important for increasing $Q^2$, being almost independent of $t$. 
The predictions of the ChPT model are also shown. As for electroproduction (Fig.~\ref{fig:hight}), the ChPT results are, in general, several orders of magnitude larger than those of the ReChi model.

\begin{figure*}[htbp]
  \centering  
      \includegraphics[width=.45\textwidth,angle=270]{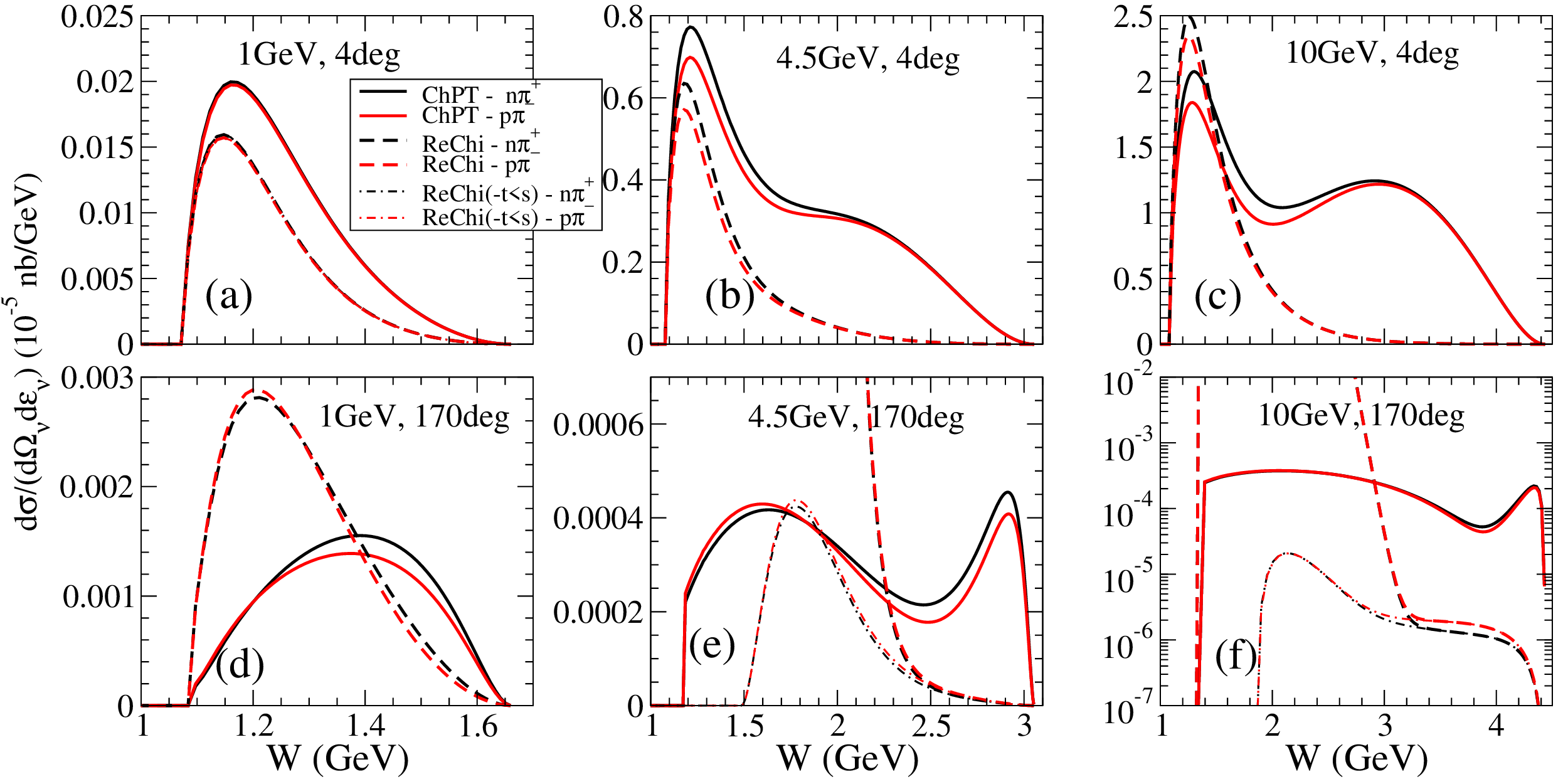}
  \caption{(Color online) As in Fig.~\ref{fig:bgs-electrons} but for WNC neutrino-induced charged-pion production.}
  \label{fig:bgs-WNC}
\end{figure*}

\begin{figure*}[htbp]
  \centering  
      \includegraphics[width=.45\textwidth,angle=270]{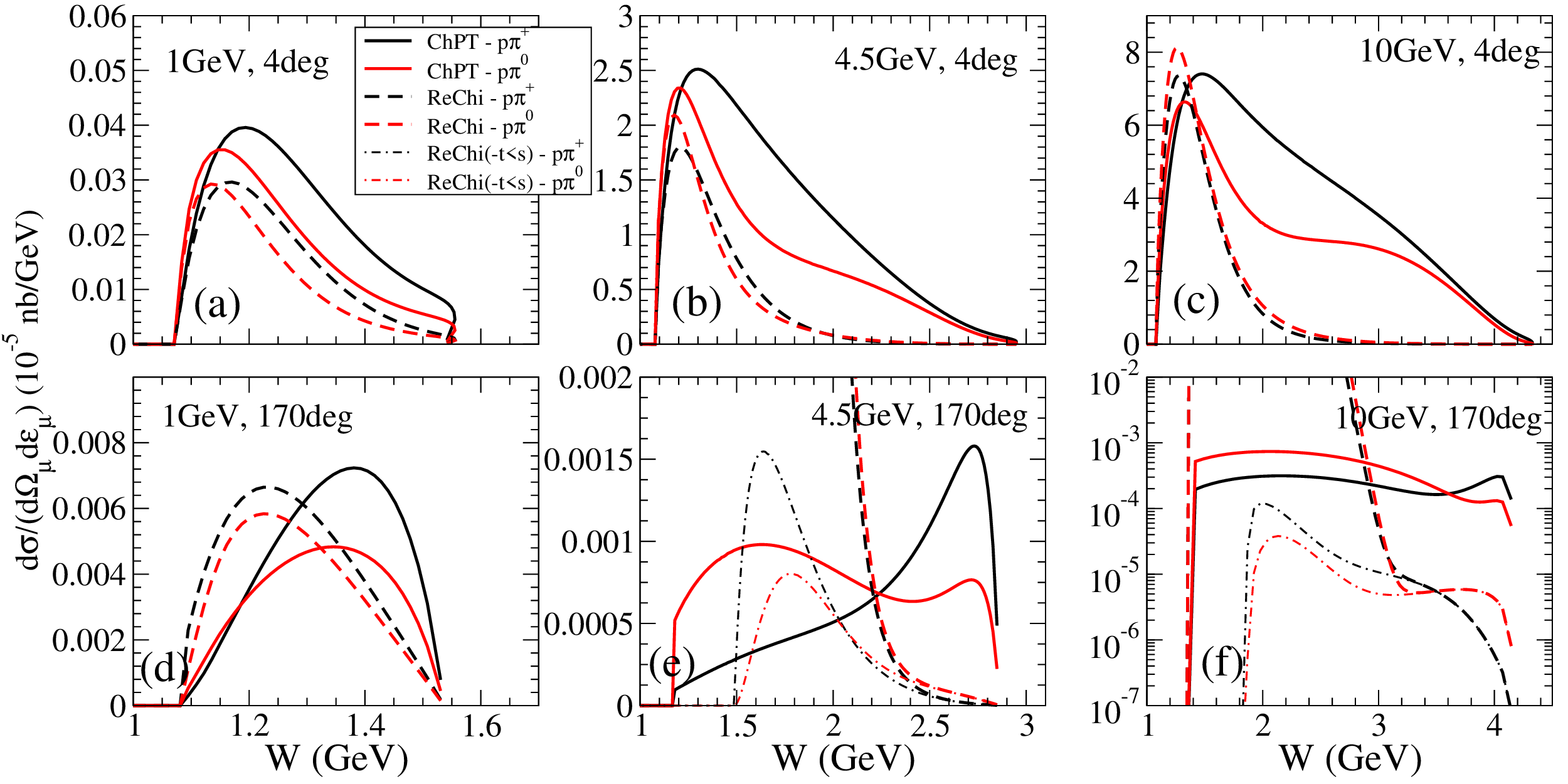}
  \caption{(Color online) As in Fig.~\ref{fig:bgs-electrons} but for CC neutrino-induced charged- and neutral-pion production. The results of the channel $n(\nu_\mu^-,\mu^-\pi^+)n$ (not shown here) are similar to the $p(\nu_\mu^-,\mu^-\pi^+)p$ ones.}
  \label{fig:bgs-CC}
\end{figure*}

To end this section, the ReChi model is compared with the ChPT-background in Fig.~\ref{fig:bgs-WNC} for WNC neutrino-induced charged-pion production and in Fig.~\ref{fig:bgs-CC} for CC neutrino-induced charged- and neutral-pion production. 
The value $\Lambda_\infty^A=7.20$ GeV is used.
The discussion of the results is similar to that of Fig.~\ref{fig:bgs-electrons}. Here, we only comment on the main differences.
At backward scattering angles and at low $W$ [panel (d)], the ReChi cross sections for neutrinos present a different behavior compared to the ChPT ones. This is due to the effect of the large parameter $\Lambda_\infty^A$ in the axial transition form factor $G_A[Q^2,s(u)]$ whose contribution is especially important at high $Q^2$ (backward lepton angles). 
Also, neutrino cross sections, contrary to electron ones, go to zero when $\varepsilon_f\rightarrow0$ (or, equivalently, for increasing $W$ values). 
This is a purely kinematic effect: neutrino cross sections are proportional to $\sigma\sim \varepsilon_f^2/M_{W,Z}^4$ while for electron cross sections one has $\sigma\sim \varepsilon_f^2/Q^4$.

\section{From low to high invariant masses: Hybrid model}\label{Hybrid-model}

In this section, we combine the low-energy model presented in Sec.~\ref{Low-energy-model} and the high-energy model presented in Sec.~\ref{High-energy-model} into a hybrid model, that can be applied over the entire $W$ region of interest for present and future accelerator-based neutrino experiments. 
Similar approaches have previously been proposed in the literature in different contexts~\cite{Barbour78,Chiang03,Jun10,Nys16}. 
In particular, the so-called Regge-plus-resonance (RPR) model developed by the Ghent group was used with remarkable success in photoproduction of strange hadrons~\cite{Corthals06,DeCruz12a,DeCruz12b}. 
However, due to the important differences between the RPR model and the present approach, we refer to the latter as {\it Hybrid model}.
Among these differences, we point out the fact that we do not apply the Regge-based model in the low-$W$ region.

The first step towards the Hybrid model is to regularize the high-energy behavior of the resonance amplitudes. 
Then, we introduce a phenomenological transition function to move from the low-energy model of Sec.~\ref{Low-energy-model} to the ReChi model of Sec.~\ref{High-energy-model}.

\subsection{Resonance cut-off form factors}\label{Res-FF}

In order to extend the low-energy model to higher invariant masses ($W>1.4$ GeV) it is mandatory to regularize the unphysical behavior of the resonance tree-level amplitudes in the kinematic regions far from the peak of the resonances, $s\approx M_R^2$.  
We do that by introducing phenomenological cut-off form factors. 

Since we are including $s$- and $u$-channel amplitudes for all resonances, the form factors should also depend on $s$ and $u$ variables. We choose the following cut-off form factor 
\ba
  F(s,u) = F(s) + F(u) - F(s) F(u)\,,\label{eq:Fsu}
\ea
which multiplies both the $s$- and $u$-channel amplitudes~\cite{Cesar06,Davidson01a,Davidson01b}. 
$F(s)$ is given by a combination of a Gaussian and a dipole form factor~\cite{Vrancx11}
\ba
  F(s) = \exp\left(\frac{- (s - M_R^2)^2}{\lambda_{R}^4 } \right)\ \frac{\lambda_R^4}{(s - M_R^2)^2 + \lambda_R^4}\,,\non\\\label{eq:F-gauss-dipole}
\ea
where $\lambda_R$ is the cut-off parameter. The same expression holds for $F(u)$ by changing $s$ by $u$.
The values $\lambda_{R_3} = 800$ MeV for spin-3/2 resonances and $\lambda_{R_1} = 1200$ MeV for spin-1/2 resonances provide the desired effect, that is,
i) avoiding the unphysical behavior of the resonances for all possible kinematics, and 
ii) keeping the original response under favorable kinematics, i.e. $s\approx M_R^2$. 
Higher spin resonances show the divergent behavior faster (at lower $s$ values) than resonances with lower spin~\cite{Vrancx11}, this explains the need of a harder form factor for spin-3/2 than for spin-1/2 resonances.

\begin{figure}[htbp]
  \centering  
      \includegraphics[width=.35\textwidth,angle=270]{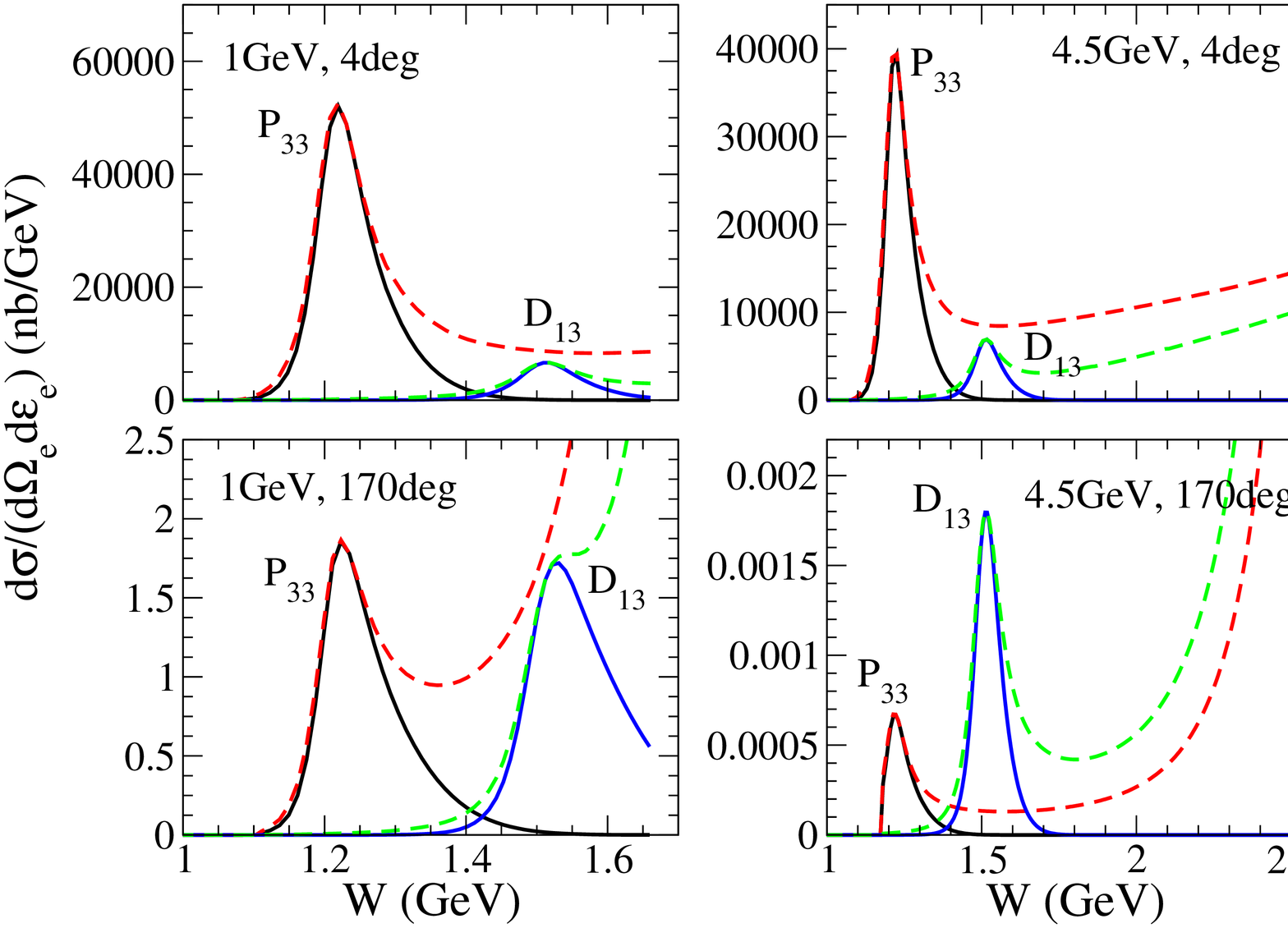}
  \caption{(Color online) Effect of the cut-off form factor on the spin-3/2 resonances. 
  The double differential cross sections for the reaction $p(e,e'\pi^+)n$ are represented as a function of the invariant mass $W$ at four different kinematics: forward ($\theta_e=4$ deg) and backward ($\theta_e=170$ deg) electron scattering angles, and incident electron energies of 1 and 4.5 GeV.
  The solid (dashed) lines are the results with (without) form factors.}
  \label{fig:FF1}
\end{figure}

\begin{figure}[htbp]
  \centering  
      \includegraphics[width=.35\textwidth,angle=270]{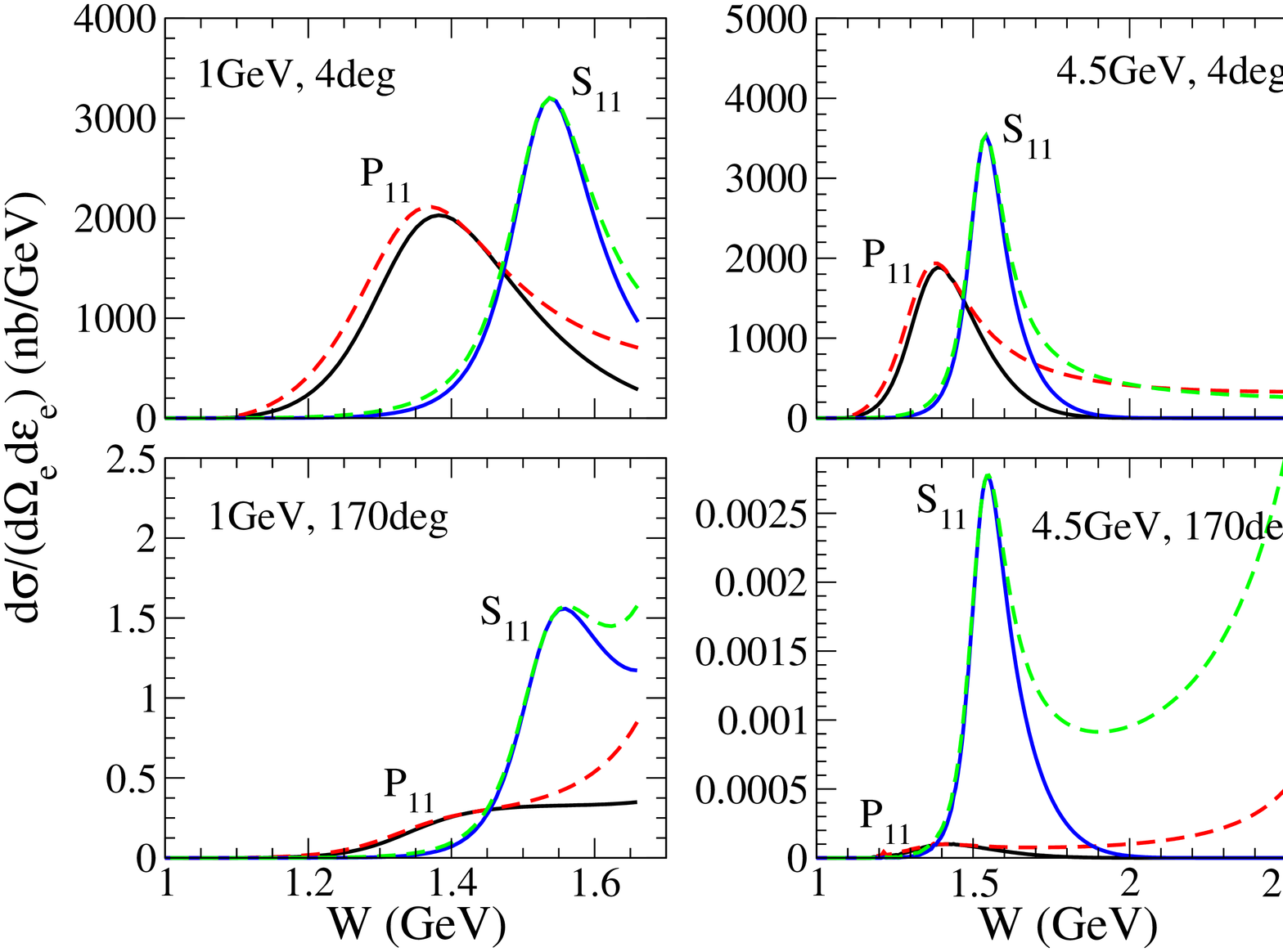}
  \caption{(Color online) As in Fig.~\ref{fig:FF1} but for spin-1/2 resonances.}
  \label{fig:FF2}
\end{figure}

The effect of these form factors in the resonances $P_{33}$ and $D_{13}$ ($P_{11}$ and $S_{11}$) is illustrated in Fig.~\ref{fig:FF1} (Fig.~\ref{fig:FF2}). 
We have represented the double differential cross sections $d\sigma/(d\Omega_e d\varepsilon_e)$ for some of the kinematic situations explained in Fig.~\ref{fig:Q2-vs-W} and for the reaction $p(e,e'\pi^+)n$ (the same discussion applies to the other reaction channels).
It is clear from the results in Figs.~\ref{fig:FF1} and \ref{fig:FF2} that the predictions without the cut-off form factors present an unphysical behavior for the $W$ region beyond the resonance peaks. 
The pathological behavior is more pronounced at backward scattering angles (higher $Q^2$), when the cross sections are much smaller, than at forward scattering angles. 
The incorporation of the cut-off form factors eliminates these tails by constraining the resonance responses to a symmetric area defined by the Gaussian-dipole shape of the form factors.

\subsection{Hybrid model}\label{sub-Hybrid-model}

The Hybrid model is constructed as follows.
We use the hadronic current operator as described in Sec.~\ref{Low-energy-model}, i.e., the ChPT background and the $s$- and $u$-channel resonance diagrams are added coherently.
The high-energy behavior of the resonances is regularized by the resonance cut-off form factors described in Sec.~\ref{Res-FF}.
In the Hybrid model, the current operator of the non-resonant contributions (${\cal O}_{ChPT}$, see Eq.~\ref{eq:O-ChPT}) is replaced by a linear combination of ${\cal O}_{ChPT}$ and the current operator of the ReChi model (${\cal O}_{ReChi}$, see Eqs.~\ref{eq:OReChiV} and \ref{eq:OReChiA}). 
This new current operator, denoted as $\widetilde{\cal O}$, is given by the transition function~\footnote{This transition function was previously used in Ref.~\cite{Gonzalez-Jimenez14b} to build the SuSAv2 model.}
\ba
  \widetilde{\cal O} =  \cos^2\phi(W) {\cal O}_{ChPT} + \sin^2\phi(W) {\cal O}_{ReChi}\,,\label{eq:wO} 
\ea
where $\phi(W)$ is a $W$-dependent function given by
\ba
  \phi(W) = \frac{\pi}{2}\left(1- \frac{1}{ 1 + \exp\left[\frac{W-W_0}{L}\right] }\right)\,.
\ea
$W_0$ and $L$ are two parameters setting the center and the width of the transition, respectively.
For instance, at $W=W_0$, one has $\widetilde{\cal O} = 1/2{\cal O}_{ChPT} + 1/2 {\cal O}_{ReChi}$ while,  
at $W=W_0+L$ and $W=W_0-L$, one has $\widetilde{\cal O}\approx{\cal O}_{ReChi}$ and $\widetilde{\cal O}\approx{\cal O}_{ChPT}$, respectively.
 
It has been mentioned along this work that the predictions from the ChPT-background model can be considered to be reliable for $W<1.4$ GeV. 
On the other hand, we have shown that the ReChi model works reasonably well under the conditions $-t/s<1$ and $W>2$ GeV.
With this in hand, we set the center of the transition at $W_0=1.7$ GeV and use a narrow transition width $L=100$ MeV which allow us to obtain a fast transition from the ChPT background to the ReChi one.

In summary, for $W<1.4$ GeV the Hybrid model is basically identical to the low energy model of Sec.~\ref{Low-energy-model}: ChPT-background terms plus resonances.
For $W>2$ GeV, the contributions from the resonances considered in this work are very small (see Figs.~\ref{fig:FF1} and \ref{fig:FF2}), therefore, it is safe to say that for $W>2$ GeV our model contains only Regge-background contributions. 
In this way, only in the transition region $1.4<W<2$ GeV, the nucleon resonances coexist with the background contributions from the ReChi model. 

\begin{figure*}[htbp]
  \centering  
      \includegraphics[width=.45\textwidth,angle=270]{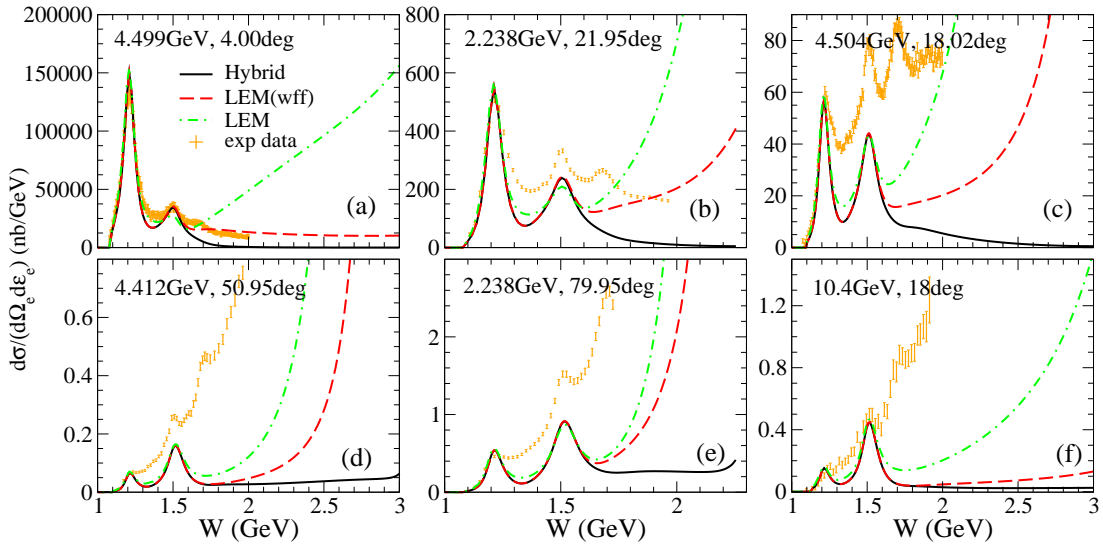}
  \caption{(Color online) Inclusive $e-p$ double differential cross section. 
  We compare the predictions from LEM (dashed-dotted-green line), LEM(wff) (dashed-red line), and Hybrid model (solid-black line) with experimental data (see text for details). 
  Data is taken from the Jefferson Lab database~\cite{JLab-database}.}
  \label{fig:eeprime}
\end{figure*}

In Fig.~\ref{fig:eeprime}, we compare our predictions with inclusive electron-proton scattering data for several values of the electron scattering angle and the incoming energy. In the inclusive process, only the scattered electron is detected. 
Therefore, since we are modeling the one-pion production channel only, we expect to underestimate the inclusive data, which include reaction channels beyond one-pion production. 
Hence, our goal is not to fit the inclusive data but to analyze the different ingredients of the model, the experimental data being an upper bound.
For this purpose, we have represented the results from three approaches:
\begin{itemize}
 \item LEM: Low-energy model presented in Sec.~\ref{Low-energy-model}. 
 \item LEM(wff): The same as LEM but with the resonance cut-off form factors of Sec.~\ref{Res-FF}. 
 \item Hybrid model: The same as LEM(wff) but using the current operator $\widetilde{\cal O}$ of Eq.~\ref{eq:wO} instead of ${\cal O}_{ChPT}$. We used the value $\Lambda_\infty^A=7.20$ GeV in our calculations. 
\end{itemize}

As expected, the three models provide basically the same results for low invariant masses $W<1.3$ GeV. 
Beyond $W>1.5$ GeV, LEM overshoots the data and makes the need of the cut-off form factors in the resonances evident. 
LEM(wff) and Hybrid model are exactly the same until $W\approx1.6$ GeV. For larger $W$, we observe huge differences between them.
We want to stress that, in the high-$W$ region, the Hybrid model provides the right magnitude of the one-pion production cross section. 
It is therefore evident that the LEM(wff) should not be used beyond $W=2$ GeV.
This is particularly obvious from the results in panel (a) and (b) where the LEM(wff) even overpredicts the inclusive data. 
Finally, it should be mentioned that in the results of the Hybrid model, the contribution from the $p(e,e'\pi^0)p$ channel is absent in (and only in) the reggeized background. In spite of that, since the magnitude of the neutral-pion electroproduction cross section is, in general, similar or smaller than the charged-pion electroproduction one~\cite{CLAS-pi014,DESY-pi075}, the present discussion is not affected.

\begin{figure*}[htbp]
  \centering
      \includegraphics[width=.45\textwidth,angle=0]{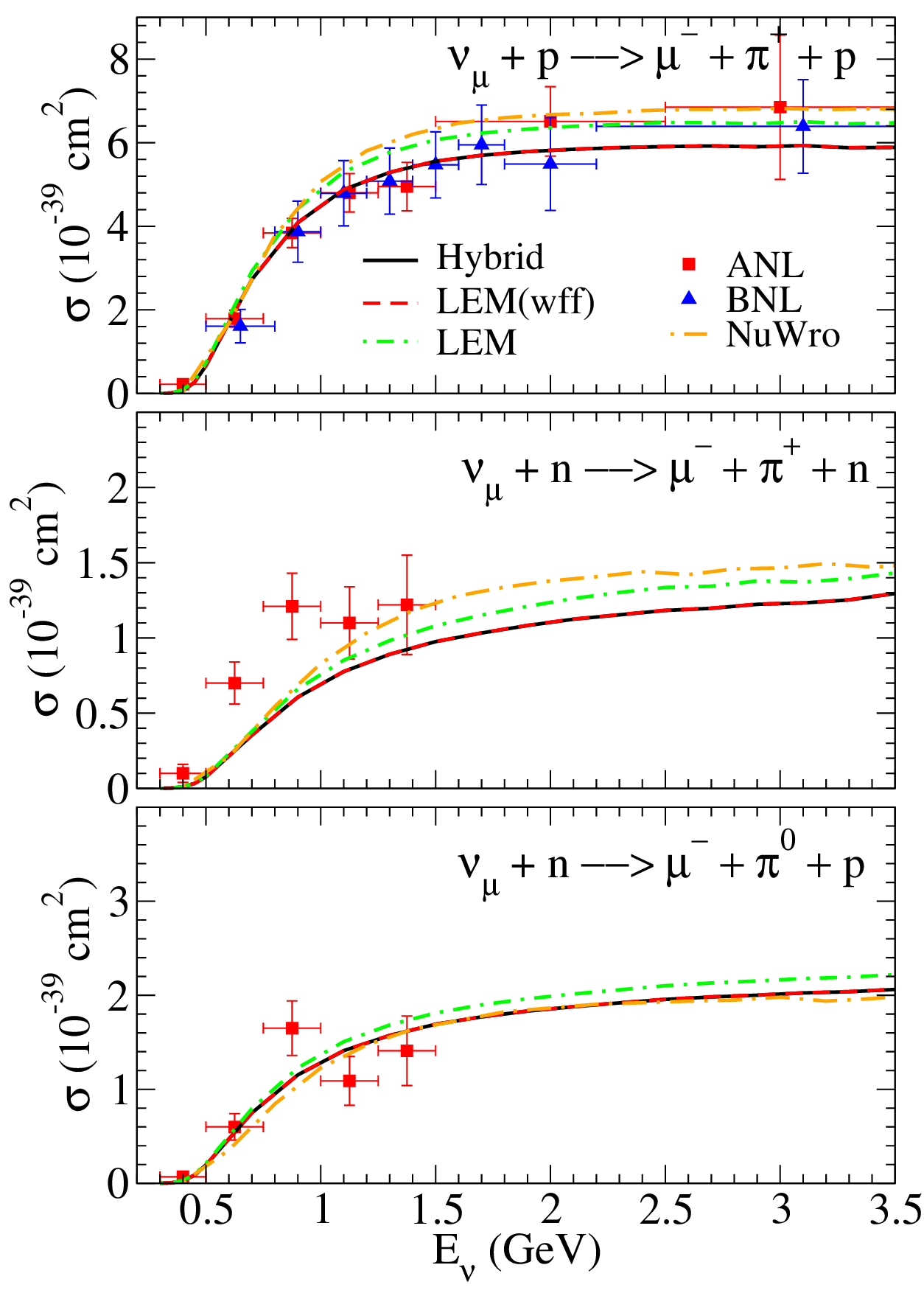}  
      \includegraphics[width=.45\textwidth,angle=0]{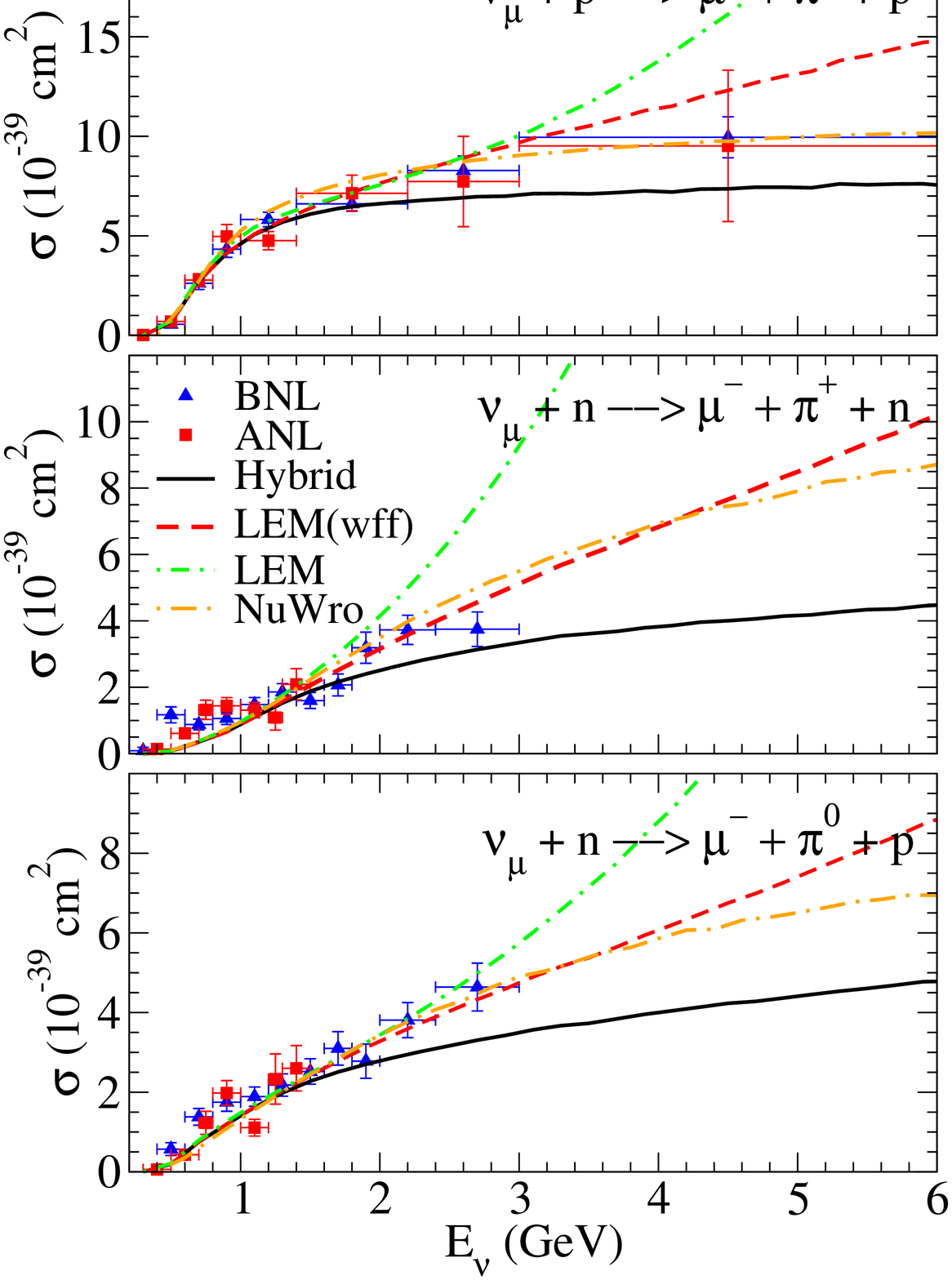}
  \caption{(Color online) (Left panels) Total cross section as a function of the neutrino energy for exclusive CC neutrino-induced one-pion production. The model predictions are compared with data from Refs.~\cite{Wilkinson14,Rodrigues16}. 
  A cut in the invariant mass $W<1.4$ GeV is applied to both model and data. 
  (Right panels) As in left panels but without the kinematic cut.}
  \label{fig:CC}
\end{figure*}

In Fig.~\ref{fig:CC}, we show the SPP total cross section for the three possible reaction channels in the neutrino-induced CC interaction. The three models described above are compared with the recent reanalysis of the BNL and ANL data~\cite{CC-ANL82_long,CC-BNL86_long}. 
We have not considered deuteron effects~\cite{Alvarez-Ruso99,Wu15} in this work.
In the left panels, a cut in the invariant mass $W<1.4$ GeV is applied and, as expected, the LEM(wff) and the Hybrid model coincide.
The effect of the cut-off form factor in the Delta resonance, which is the only resonance playing a significant role at $W<1.4$ GeV, produces a reduction of the cross sections of approximately $5-15\%$, depending on the neutrino energy and the reaction channel.
In general, the three models reproduce the $p(\nu_\mu,\mu^-\pi^+)p$ and $n(\nu_\mu,\mu^-\pi^0)p$ data well, but they underestimate the $n(\nu_\mu,\mu^-\pi^+)n$ data. This is a well-known problem of the low-energy model~\cite{Hernandez07}, which may be related to the role of the cross-Delta-pole diagram and deuteron effects~\cite{Alvarez-Ruso16,Wu15}.

There is no cut in the right panels in Fig.~\ref{fig:CC}, so higher invariant masses contribute to the cross sections.
We will focus on the comparison between models and data in the energy region $\varepsilon_i>2$ GeV, where the predictions of the three models are clearly different.
In this region, the LEM overestimates the data while the Hybrid model underestimates them. 
LEM(wff) lays in between the others and seems to be the one in better agreement with data.
However, LEM(wff) cannot be consider realistic at these energies since it contains unphysical strength from the $W>2$ GeV region (see, for instance, panel (b) of Fig.~\ref{fig:CC}). Therefore, we interpret this apparent agreement as a mere coincidence.

\begin{figure}[htbp]
  \centering  
      \includegraphics[width=.35\textwidth,angle=270]{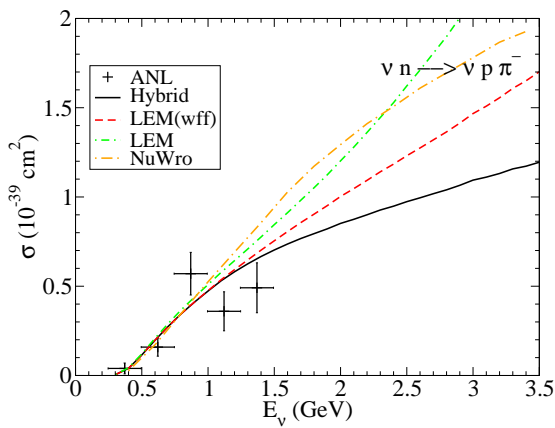}
  \caption{(Color online) Total cross section as a function of the neutrino energy for the WNC reaction $n(\nu,\nu\pi^-)p$. 
  The model is compared to ANL data~\cite{WNC-ANL80}.}
  \label{fig:WNC}
\end{figure}

Finally, in Fig.~\ref{fig:WNC} we compare the WNC $n(\nu,\nu\pi^-)p$ total cross section data~\cite{WNC-ANL80} with our predictions.
In this case, only low energy data ($\varepsilon_i<1.5$ GeV) are available and Hybrid model is the one with the better agreement with data.

In Figs.~\ref{fig:CC} and \ref{fig:WNC}, we have included the predictions of NuWro (see previous discussion in Sect.~\ref{High-energy-model}). 
% In NuWro, the SPP channel on the nucleon is defined by selecting those events with only one pion and one nucleon in the final state. 
%
In NuWro, the reaction mechanisms in the region $W<1.6$ GeV are the Delta resonance pole and an effective background extrapolated from the DIS contribution. 
For $W>1.6$ GeV the predictions are based on the DIS formalism~\cite{Bodek02} and the PYTHIA 6 hadronization routines~\cite{Pythia6}. 
A smooth transition between the resonance and DIS regions is performed in the region $1.4 < W < 1.6$ GeV (see Ref.~\cite{Sobczyk05} for details).
The parameters of the axial form factor of the Delta resonance,  $C_5^A(0)=1.19$ and $M_A=0.94$ GeV, were fitted to reproduce the original BNL and ANL data~\cite{Graczyk09}.
This can partially explain why the NuWro predictions are systematically larger than the Hybrid model.

\begin{figure}[htbp]
  \centering  
      \includegraphics[width=.45\textwidth,angle=0]{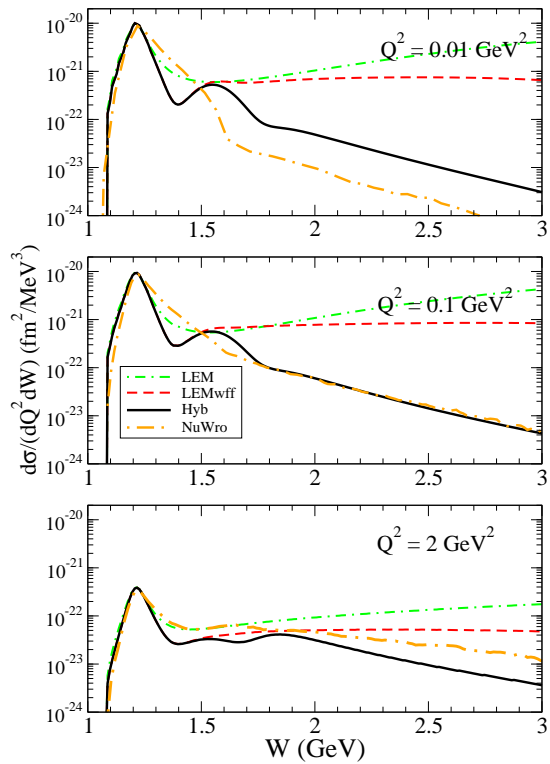}
  \caption{(Color online) Different model predictions for the differential cross section $d\sigma/(dQ^2dW)$, for the channel $p(\nu_\mu,\mu^-\pi^+)p$. The incoming neutrino energy is fixed to $E_\nu=10$ GeV. }
  \label{fig:ds-dQ2dW}
\end{figure}

Finally, in Fig.~\ref{fig:ds-dQ2dW} we compare the predictions of NuWro and the Hybrid model as a function of $W$ for three values of $Q^2$. The predictions of the LEM and the LEM(wff) are also shown as reference.
This comparison helps to further understand the differences between the two models.
First we analyze the resonance region. In NuWro, only the Delta is present while, in the Hybrid model, the peak from higher mass resonances appears around $W\approx1.5$ GeV. Also, the width of the Delta is considerably larger in NuWro than in the other models.
In the high energy region ($W>2$ GeV), both NuWro and the Hybrid model predict a rapidly decreasing behavior for increasing $W$ values. 
Still, the predictions show a different $Q^2$ behavior.
% For increasing $Q^2$, the predictions of NuWro are considerable larger than those from the Hybrid model while 
%
% The predictions of the Hybrid model are presented for the two situations that were analyzed in Sect.~\ref{High-energy-model}, i.e., $\Lambda_\infty^A=7.2$ and 2.194 GeV, referred as Hyb(7.2) and Hyb(2.2) in the figure. 
The bump in the cross section, appearing at $W\approx1.8$ GeV for the Hybrid model and at $W\approx1.6$ GeV in NuWro, are caused by the modeling of the transition region. 
The incorporation of higher mass resonances may push the reliability of the low-energy model further in $W$, what would help to solve this problem. 
Further studies in the modeling of both the resonance and the high energy regions are needed.

\section{Conclusions and outlook}\label{Conclusions}

We have developed a model for electroweak SPP off the nucleon that is applicable for invariant masses from the pion-production threshold to the high-$W$ limit. 
Within our approach, the reaction is modeled at the amplitude level, this allows one to make predictions for fully exclusive reactions, with information on the angular distributions of all outgoing particles.

Our starting point is the electroweak SPP low-energy model summarized in Sec.~\ref{Low-energy-model}. It includes contributions from nucleon resonances and background terms derived from the ChPT Lagrangian of the $\pi N$ system (Appendix~\ref{ChPT-expressions}).
We have shown that this model works well for small invariant masses, $W\lesssim1.4$ GeV, but fails beyond. 

In Sec.~\ref{High-energy-model}, this low energy-model was extended to the high-energy region using Regge phenomenology. By the ``standard procedure''~\cite{Guidal97} of replacing the $t$-channel Feynman propagators with the corresponding Regge trajectories, we reggeized the ChPT-background contributions. 
The result is a high-energy model for electroweak SPP (ReChi model) whose validity, by construction, is restricted to the region $W>2$ GeV and forward $\theta_\pi^*$ scattering angles.
Since the forward $\theta_\pi^*$ scattering region strongly dominates the one-pion production cross section, one expects that the predictions of the ReChi model underestimate only slightly the $\theta_\pi^*$-integrated cross sections.

In a scheme of strong degeneracy, we reggeized the EM current considering only the $\pi(140)/b_1(1235)$-Regge trajectory.
To obey CVC, the $CTv$ and the $s(u)$-channel Born terms were included, with phenomenological transition form factors as in Refs.~\cite{Kaskulov10a,Vrancx14a}. 
We found an acceptable agreement between the ReChi model and charged-pion electroproduction data in the region $0<-t<4$ GeV$^2$, $W\approx2$ GeV and $Q^2<4$ GeV$^2$. 
In spite of the dominance of the degenerate $\pi(140)/b_1(1235)$ Regge trajectory in the vector current, more sophisticated models containing additional trajectories such as $\rho(770)/a_2(1320)$ and $a_1(1260)$~\cite{Guidal97,Kaskulov10a}, as well as other ingredients beyond the tree-level amplitudes~\cite{Laget06,Laget10}, may be preferable.
However, the main goal of this work is to make predictions in the weak sector. For that, one also needs to model the axial current, which is absent in the above mentioned EM models.

The vector current for neutrino interactions was obtained from the EM one by isospin rotation. 
The real challenge is the reggeization of the axial current, which has no analogous counterpart in electron scattering, and experimental information is scarce.
For that, we have reinterpreted the $PP$ and the axial part of the $CT$ as effective $\rho$-exchange $t$-channel diagrams. 
This allowed us to reggeize the axial current by considering the $\rho$ exchange as the main Regge trajectory.
We have compared the ReChi model with total cross section data~\cite{Allen86} for the CC reactions 
$\bar\nu p\rightarrow\mu^+\pi^- p$ and $\nu p\rightarrow\mu^-\pi^+ p$ in the energy range between 10 and 90 GeV. The data does not include contributions from the $W<2$ GeV region, so only residual effects from the resonance region are expected.
We showed that, to reproduce the magnitude of the experimental data, we needed a large value of the axial transition form factor parameter, $\Lambda_\infty^A$, which is difficult to interpret. 
We believe that other ingredients not considered in the ReChi model may still play an important role in the one-pion production cross section at high invariant masses. In particular, the contribution of other meson trajectories and the modeling of the backward $\theta_\pi^*$ scattering region need to be further investigated. 
Also, we may be missing strength from the high-$Q^2$ sector of the cross section. At high $Q^2$ and high $W$, one enters in the DIS region, where the direct interaction with partons makes the use of hadrons as effective degrees of freedom rather questionable.

We have included predictions of the NuWro Monte Carlo event generator. 
The NuWro predictions for the antineutrino reaction are a factor $\sim2$ larger than those of the neutrino counterpart. 
This result is consistent with the DIS treatment of the problem that is implemented in NuWro, but it does not seem to be supported by data. 
More experimental data would certainly help to clarify which are the main reaction mechanisms playing a role in SPP in the region $W>2$ GeV.

Finally, in Sec.~\ref{Hybrid-model} we have proposed a phenomenological way of combining the low- and high-energy models into a Hybrid model that can be applied in the entire region of interest for accelerator-based neutrino oscillation experiments. For that, it was also necessary to regularize the $W>M_R$ behavior of the $s$ and $u$ channels of the resonances, this was done by using phenomenological form factors (Sec.~\ref{Res-FF}).
The Hybrid model was compared with the neutrino-induced SPP total cross section data from the BNL and ANL experiments, as well as with the NuWro predictions.
We showed that the low-energy model and the Hybrid model agree well up to neutrino energies around $\sim1$ GeV. 
Beyond that, the contributions from the region $W>1.4$ GeV start to be important, and the low-energy model fails. 
The resonance cutoff form factors and the reggeization of the ChPT background allowed us to provide reliable predictions in this region. 
NuWro predicts systematically larger cross sections than the Hybrid model. 
This is due to the fact that the Delta and the high-$W$ contributions in NuWro are larger than in the Hybrid model.

In order to perform predictions of neutrino cross sections, in particular, in the case of neutrino-nucleus interaction, one needs computationally efficient models. 
The low-energy model employed in this work has been used by several collaborations~\cite{Hernandez07,Lalakulich10,Sobczyk13,Rafi16}. 
Our proposal for extending this model to higher invariant masses is technically and formally straightforward, and does not involve any additional cost from a computational point of view. 
Work is in progress to apply this model to the case of neutrino-induced one-pion production off nuclei. 
A preliminary work in which the nucleons are described within a relativistic mean-field model was presented in Ref.~\cite{Gonzalez-Jimenez16a}. Only the low-energy model was considered in that reference.

We end this paper by providing some final remarks about possible future improvements.

The ReChi model does not describe neutral-pion production induced by neutral-current (EM and WNC) interactions because the meson-exchange diagrams needed to reggeize the ChPT model do not contribute to the amplitude: the isospin factors of the $PF$, $CT$ and $PP$ diagrams are zero (Table~\ref{Iso-coef-EM}). This is also explained by the fact that the vertices $\rho^0\to\pi^0\pi^0$ and $a_1^0(1260)\to\rho^0\pi^0$ are forbidden by isospin symmetry.
In Refs.~\cite{Laget11,Kaskulov10b}, the neutral-pion electroproduction data~\cite{CLAS-pi014,DESY-pi075} are reasonably well described within a Regge framework. 
However, ingredients beyond the ChPT diagrams considered in this work are needed and, obviously, the description of the axial current is missing.
The recent publication by the MINERvA collaboration~\cite{MINERvAdiff16} of ``evidence for WNC-diffractive $\pi^0$ production from hydrogen'' points out the urgency of theoretical predictions for this process. 

Other higher mass resonances, beyond the $D_{13}(1520)$ and $S_{11}(1535)$ considered in this work, still play an important role and should be included in order to reproduce the broad peak in the cross section observed around $W\approx1.7$ GeV (Fig.~\ref{fig:eeprime}).

The ReChi and Hybrid model presented here are flexible and there is still room for improvements. 
A proper fit and fine tunning of the parameters and form factors of the model as well as more investigations regarding the modeling of the axial current in the high-energy model are requiered, in particular, if more neutrino-induced SPP data is available.
In this sense, this work should be understood as a first approach to the problem, further analyses and improvements are desired and expected in forthcoming publications.

\section*{Acknowledgements}

This work was supported by the Interuniversity Attraction Poles Programme initiated by the Belgian Science Policy Office (BriX network P7/12) and the Research Foundation Flanders (FWO-Flanders). The computational resources (Stevin Supercomputer Infrastructure) and services used in this work were provided by Ghent University, the Hercules Foundation and the Flemish Government.
J.N. was supported as an `FWO-aspirant'.
V.P. acknowledges the support by the National Science Foundation under Grant No. PHY-1352106.
K.N. was partially supported by the NCN grant UMO-2014/14/M/ST2/00850.
R.G.J. wants to thank L. Alvarez-Ruso for some clarifications about the unitarization of the low-energy model, J. Ryckebusch for inspiring comments about Regge theory, C. Colle for assisting with the codes, and W. Cosyn for helping with the ChPT. 
We also thank W. Cosyn and J. Sobczyk for a careful reading of this manuscript.

\appendix

\section{Pion-nucleon system in ChPT: Background contributions}\label{ChPT-expressions}

The ChPT Lagrangian for the pion-nucleon system provides all the necessary vertices for computing the background Feynman diagrams considered in this model. In this appendix we present a detailed derivation of those vertices. 
We also show how to identify the vector and axial currents of the electroweak process. 
A different derivation of the vector and axial currents of the pion-nucleon system based on the transformation properties of the fields can be found in Ref.~\cite{Hernandez07}.
We closely follow the procedure and convention of Ref.~\cite{Scherer12}. 
Some of the expressions presented here can be found in Refs.~\cite{Scherer12,Hilt13},
we refer the reader to those references for further details. 

ChPT applied to the pion-nucleon system with coupling to external fields provides the following effective Lagrangian:
\ba
  {\cal L}_{eff} = {\cal L}_{\pi N} + {\cal L}_\pi\,,
\ea
where
\ba
  {\cal L}_{\pi N} &=& \nPsib \left(i\Dslash - M + ig_A\gamma^\mu\gamma^5{\cal A}_\mu \right)\nPsi\,,\\
  {\cal L}_{\pi} &=& \frac{f_\pi^2}{4}\text{Tr}\left[ D_\mu U (D^\mu U)^\dag \right]\,.  
\ea
The covariant derivative $D_\mu$ is defined as
\ba
  D_\mu = \pd + {\cal V}_\mu - iv_\mu^{(s)}\,,
\ea
and
\ba
  {\cal V}_\mu = \frac{1}{2}\Big[u^\dag(\pd-ir_\mu)u + u(\pd-il_\mu)u^\dag\Big]\,,\\
  {\cal A}_\mu = \frac{1}{2}\Big[u^\dag(\pd-ir_\mu)u - u(\pd-il_\mu)u^\dag\Big]\,.
\ea
The matrix $u$ is the square root of the matrix $U$ containing the pion fields ($\vec\phi$):
\ba
U = \exp\left(i\frac{\vec\ntau\cdot\vphi}{f_\pi}\right)\,,\,\,\,\,\, u = \exp\left(i\frac{\vec\ntau\cdot\vphi}{2f_\pi}\right)\,.
\ea

The fields $r_\mu$, $l_\mu$ and $v_\mu^{(s)}$ provide the coupling to the external boson fields:
\ba
  r_\mu &=& -e\frac{\ntau_z}{2}A_\mu + \frac{g}{2\cos\theta_W}\sin^2\theta_W \ntau_z Z_\mu\,,\\
  l_\mu &=& -e\frac{\ntau_z}{2}A_\mu + \frac{-g\ \Cab}{\sqrt2}\left(\ntau_+W^+_\mu + \ntau_-W^-_\mu\right) \non\\
      &+& \frac{-g}{2\cos\theta_W}(1-\sin^2\theta_W)\ntau_z Z_\mu\,,\\
  v_\mu^{(s)} &=& -\frac{e}{2}A_\mu + \frac{g}{2\cos\theta_W}\sin^2\theta_W Z_\mu\,,
\ea
where $A_\mu$, $W_\mu^{+/-}$ and $Z_\mu$ are the fields of the photon, W$^{+/-}$ boson and Z boson, respectively.
  
\begin{widetext}
An expansion in terms of the pion decay constant $f_\pi=93$ MeV, where only the interaction terms of the Lagrangian which contribute to one-pion production at first order in $f_\pi$ are kept, results in the following effective Lagrangian:
\ba
{\cal L}_{eff} &=& {\cal L}_{\pi NN} + {\cal L}_{\pi\pi NN}\non\\
	&+& {\cal L}_{\gamma NN} + {\cal L}_{\gamma\pi\pi} + {\cal L}_{\gamma\pi NN} \non\\
	&+& {\cal L}_{W NN} + {\cal L}_{W\pi} + {\cal L}_{W\pi\pi} + {\cal L}_{W\pi NN,V} + {\cal L}_{W\pi NN,A}\non\\
	&+& {\cal L}_{Z NN} + {\cal L}_{Z\pi} + {\cal L}_{Z\pi\pi} + {\cal L}_{Z\pi NN,V} + {\cal L}_{Z\pi NN,A}\,.\label{eq:Leff}
\ea  
Each term of the effective Lagrangian in Eq.~\ref{eq:Leff} provides a different vertex function. 
According to this, the possible Feynman diagrams contributing to the one-pion production amplitude (at first order in $1/f_\pi$) are shown in Fig.~\ref{fig:nres-diagrams}.

In what follows, we present the explicit expressions for each term of the Lagrangian in Eq.~\ref{eq:Leff}. 
Using these expressions together with the appropriate Feynman rules, it is straightforward to obtain the background contributions to the hadronic current of Eq.~\ref{eq:Jhad}.
We define the physical fields of the pion as $\phi^0 \equiv \phi_z$, $\phi^+ \equiv \frac{1}{\sqrt2}(\phi_x-i\phi_y)$ and $\phi^- \equiv \frac{1}{\sqrt2}(\phi_x+i\phi_y)$.
The convention is such that $\phi^0$ creates or annihilates a $\pi^0$, and $\phi^+(\phi^-)$ annihilates a $\pi^+(\pi^-)$ or creates a $\pi^-(\pi^+)$. 
The same convention is used for the boson fields.\\

  \paragraph{Pion-nucleon Lagrangians.}
The terms containing pion and nucleon fields only are:
\ba
{\cal L}_{\pi NN} &=& -\frac{g_A}{2f_\pi}
	 \Big[ \psib_p(\gamma^\mu\gamma^5\pd\phi^0)\psi_p 
	 - \psib_n(\gamma^\mu\gamma^5\pd\phi^0)\psi_n\non\\
	 &+& \sqrt2\ \psib_p(\gamma^\mu\gamma^5\pd\phi^+)\psi_n
	+ \sqrt2\ \psib_n(\gamma^\mu\gamma^5\pd\phi^-)\psi_p \Big],
	\label{eq:LpiNN}
\ea
\ba
{\cal L}_{\pi\pi NN} &=& -\frac{i}{4f_\pi^2} \Big[ 
	  \psib_p\gamma^\mu(\phi^-\pd\phi^+ - \phi^+\pd\phi^-)\psi_p
	- \psib_n\gamma^\mu(\phi^-\pd\phi^+ - \phi^+\pd\phi^-)\psi_n\non\\
	&+& \sqrt2\ \psib_p\gamma^\mu(\phi^+\pd\phi^0 - \phi^0\pd\phi^+)\psi_n
	+ \sqrt2\ \psib_n\gamma^\mu(\phi^0\pd\phi^- - \phi^-\pd\phi^0)\psi_p\Big]\,.
	\label{eq:LpipiNN}
\ea

  \paragraph{Couplings to external fields: photon.}
The $\gamma NN$ vertex reads:  
\ba
    {\cal L}_{\gamma NN} &=& -e\psib_p\gamma^\mu\psi_p A_\mu\non\\
			  &\rightarrow&  -e(\psib_p \hat{\Gamma}_p^\mu\psi_p 
			  + \psib_n \hat{\Gamma}_n^\mu\psi_n) A_\mu\,,\label{eq:LphNN}
\ea
with
\ba
  \hat{\Gamma}_{p,n}^\mu \equiv F_1^{p,n}\gamma^\mu + i\frac{F_2^{p,n}}{2M}\sigma^{\mu\alpha}Q_\alpha\,.\label{eq:GgNN}
\ea
In Eq.~\ref{eq:LphNN} we have replaced the point-like photon-proton-proton coupling by the vertex function which takes into account the inner structure of the nucleon. $F_{1,2}^{p,n}$ are the proton and neutron electromagnetic form factors for which we use the Galster parametrization~\cite{Galster71}.

The $\gamma \pi\pi$ and $\gamma\pi NN$ vertices are:
\ba
    {\cal L}_{\gamma\pi\pi} &=&  -ie(\phi^-\pd\phi^+ - \phi^+\pd\phi^-)A_\mu\,.\label{eq:Lphpipi}\\
  {\cal L}_{\gamma\pi NN} &=& \frac{-ieg_A}{\sqrt2f_\pi}
	(\psib_p\gamma^\mu\gamma^5\phi^+\psi_n
      - \psib_n\gamma^\mu\gamma^5\phi^-\psi_p)\ A_\mu\,. \label{eq:LphpiNN}
\ea  

We can rewrite the previous expressions of the Lagrangians as ${\cal L} = J^\mu_{EM} A_\mu$, where the photon couples to a current $J^\mu$. 
This current can be identified as the EM current of the pion-nucleon system:
\ba
  J_{EM}^\mu = J_{\gamma NN}^\mu + J_{\gamma\pi\pi}^\mu + J_{\gamma\pi NN}^\mu\,.
\ea
It is easy to show that $J_{EM}^\mu$ transforms as a vector under parity transformation.\\

  \paragraph{Couplings to external fields: W boson.}
The $W NN$ vertex reads:  
\ba
{\cal L}_{W NN} &=& \frac{-g\Cab}{2\sqrt2}\Big[ \psib_p\gamma^\mu(1-g_A\gamma^5)\psi_n W_\mu^+ 
	  + \psib_n\gamma^\mu(1-g_A\gamma^5)\psi_p W_\mu^- \Big]\non\\
	      &\rightarrow& \frac{-g\Cab}{2\sqrt2}\Big[ \psib_p(\hat{\Gamma}^\mu_V-\hat{\Gamma}^\mu_A)\psi_n W_\mu^+ 
	  + \psib_n(\hat{\Gamma}^\mu_V-\hat{\Gamma}^\mu_A)\psi_p W_\mu^- \Big]\,,\label{eq:LWNN}
\ea
with
\ba
  \hat{\Gamma}^\mu_V &\equiv& F_1^V\gamma^\mu + i\frac{F_2^V}{2M}\sigma^{\mu\alpha}Q_\alpha\,,\non\\
  \hat{\Gamma}^\mu_A &\equiv& G_A(\gamma^\mu\gamma^5 + \frac{\Qslash}{m_\pi^2-Q^2}Q^\mu\gamma^5)\,.\label{eq:GWNN}
\ea
As in Eq.~\ref{eq:LphNN}, we have replaced the point-like coupling by the vertex function which takes into account the inner structure of the nucleon.
The isovector vector form factor is $F_{1,2}^V = F_{1,2}^p-F_{1,2}^n$.
$G_A$ is the isovector axial form factor given by the usual dipole form $G_A(Q^2)=g_A/(1-Q^2/M_A^2)^2$ with $g_A=1.26$ and $M_A=1.05$ GeV.

The $W \pi$, $W \pi\pi$ vertices are:
\ba
  {\cal L}_{W\pi} &=& -\frac{g}{2}f_\pi\Cab (\pa\phi^-W^+_\mu + \pa\phi^+W^-_\mu)\,,\label{eq:LWpi}\\
  {\cal L}_{W\pi\pi} &=& i\frac{g}{2}\Cab \Big[(\phi^-\pa\phi^0 - \pa\phi^-\phi^0)W_\mu^+ 
		    + (\phi^+\pa\phi^0 - \pa\phi^+\phi^0)W_\mu^-\Big]\,.\label{eq:LWpipi}
\ea
The $W\pi NN$ vertex has a vector and an axial contributions:
\ba
  {\cal L}^A_{W\pi NN} &=& \frac{g\Cab}{2\sqrt2}\frac{-i}{\sqrt2f_\pi} \Big[\left(\sqrt2\ \psib_p\gamma^\mu\phi^0\psi_n
		      - \psib_p\gamma^\mu\phi^-\psi_p 
		      + \psib_n\gamma^\mu\phi^-\psi_n \right)W_\mu^+\non\\
		      &+& \left(\psib_p\gamma^\mu\phi^+\psi_p 
		      - \psib_n\gamma^\mu\phi^+\psi_n 
		      - \sqrt2\ \psib_n\gamma^\mu\phi^0\psi_p \right)W_\mu^- \Big]\,,\label{eq:LWpiNN_A}\\
  {\cal L}^V_{W\pi NN} &=& \frac{g\ g_A\Cab}{2\sqrt2}\frac{i}{\sqrt2f_\pi} \Big[
			\left(\sqrt2\ \psib_p\gamma^\mu\gamma^5\phi^0\psi_n 
		      - \psib_p\gamma^\mu\gamma^5\phi^-\psi_p 
		      + \psib_n\gamma^\mu\gamma^5\phi^-\psi_n \right)W_\mu^+\non\\
		      &+& \left(\psib_p\gamma^\mu\gamma^5\phi^+\psi_p 
		      - \psib_n\gamma^\mu\gamma^5\phi^+\psi_n 
		      -\sqrt2\ \psib_n\gamma^\mu\gamma^5\phi^0\psi_p \right)W_\mu^- \Big]\,.\label{eq:LWpiNN_V}
\ea

The current which couples to the W boson and that can be identified as the weak charged-current of the pion-nucleon system (${\cal L}=J^\mu_{CC} W^{\pm}_\mu$) has a vector part
\ba
  J_{CC,V}^\mu = J_{W NN,V}^\mu + J_{W\pi\pi}^\mu + J_{W\pi NN,V}^\mu\,,
\ea 
and an axial part: 
\ba
  J_{CC,A}^\mu = J_{W NN,A}^\mu + J_{W\pi}^\mu + J_{W\pi NN,A}^\mu\,,
\ea
which, under parity transformations, transforms as a vector and an axial vector, respectively. \\

  \paragraph{Couplings to external fields: Z boson.}
The $Z NN$ vertex reads:  
\ba
{\cal L}_{Z NN} &=& \frac{-g}{2\cos\theta_W}
\Big[ \psib_p\frac{1}{2}\gamma^\mu(Q^p_W - g_A\gamma^5)\psi_p Z_\mu 
  +\psib_n\frac{1}{2}\gamma^\mu(Q_W^n + g_A\gamma^5)\psi_n Z_\mu \Big]\non\\
  &\rightarrow& \frac{-g}{2\cos\theta_W}
  \Big[ \psib_p(\hat{\Gamma}^\mu_{p,V}-\hat{\Gamma}^\mu_{p,A})\psi_p Z_\mu 
	  + \psib_n(\hat{\Gamma}^\mu_{n,V}-\hat{\Gamma}^\mu_{n,A})\psi_n Z_\mu \Big]\,.
	  \label{eq:LZNN}
\ea
$Q_W^{p,n}$ represent the weak vector charge of the proton and neutron seen by the neutrino.
Neglecting higher order radiative corrections their values are given by $Q_W^p=(1-4\sin^2\theta_W)$ and $Q_W^n=-1$. 
To take into account the inner structure of the nucleon we have introduced
\ba
  \hat{\Gamma}^\mu_{p(n),V} &\equiv& \widetilde F_1^{p(n)}\gamma^\mu + i\frac{\widetilde F_2^{p(n)}}{2M}\sigma^{\mu\alpha}Q_\alpha\,,\\
  \hat{\Gamma}^\mu_{p(n),A} &\equiv& \widetilde G_A^{p(n)}(\gamma^\mu\gamma^5 + \frac{\Qslash}{m_\pi^2-Q^2}Q^\mu\gamma^5)\,.\label{eq:GZNN}
\ea
The WNC vector form factors of the proton and neutron can be related to the EM ones by
\ba
  \widetilde F_{1,2}^{p,n} =  \frac{1}{2}Q_W^pF_{1,2}^{p,n} + \frac{1}{2}Q_W^n F_{1,2}^{n,p} - \frac{1}{2}F_{1,2}^{s}\,,\label{eq:WNC-ff}
\ea
with $F_{1,2}^{s}$ the strange form factor of the nucleon. 
The WNC axial form factor can be related to the isovector axial form factor $G_A(Q^2)$ that enter in the CC interaction by
\ba
  \widetilde G_A^{p(n)} = (\pm)\frac{1}{2}G_A - \frac{1}{2}G_A^{s}\,,
\ea
with for $+$ ($-$) for proton (neutron) and $G_A^{s}$ the axial strange nucleon form factor.
The strange form factors of the nucleon have recently been studied using the parity-violating electron scattering asymmetry~\cite{Jachowicz07,Gonzalez-Jimenez13a,Gonzalez-Jimenez14a,Gonzalez-Jimenez15a,Gonzalez-Jimenez15b,Moreno15} and the proton-to-neutron ratio in WNC neutrino-nucleus quasielastic interaction~\cite{Gonzalez-Jimenez13b,Gonzalez-Jimenez13c}.
A general conclusion from these works is that we are still far from a precise determination of the strange form factor of the nucleon. Therefore, given the large uncertainties in our model from other sources, for simplicity, in this work we fix all strangeness contributions to zero.

The $Z \pi$, $Z \pi\pi$ vertices are:
\ba
{\cal L}_{Z\pi} &=& \frac{-g}{2\cos\theta_W}f_\pi \pa\phi^0 Z_\mu\,,\\ \label{eq:LZpi}
{\cal L}_{Z\pi\pi} &=& i\frac{-g}{2\cos\theta_W}(1-2\sin^2\theta_W) (\phi^-\pa\phi^+ - \phi^+\pa\phi^-)Z_\mu\,,\label{eq:LZpipi}
\ea
The $Z\pi NN$ vertex has a vector and an axial contributions:
\ba
{\cal L}^A_{Z\pi NN} &=& \frac{g}{2\cos\theta_W}\frac{i}{\sqrt2f_\pi} 
		    \left(\psib_p\gamma^\mu\phi^+\psi_n - \psib_n\gamma^\mu\phi^-\psi_p \right)Z_\mu\,,\label{eq:LZpiNN_A}\\
{\cal L}^V_{Z\pi NN} &=& \frac{-g\ g_A}{2\cos\theta_W}\frac{i(1-2\sin^2\theta_W)}{\sqrt2f_\pi} 
  \left(\psib_p\gamma^\mu\gamma^5\phi^+\psi_n - \psib_n\gamma^\mu\gamma^5\phi^-\psi_p \right)Z_\mu\,.\label{eq:LZpiNN_V}
\ea

The current which couples to the Z boson and that can be identified as the weak neutral-current of the pion-nucleon system (${\cal L}=J^\mu_{WNC}Z_\mu$) has a vector part:
\ba
  J_{WNC,V}^\mu = J_{Z NN,V}^\mu + J_{Z\pi\pi}^\mu + J_{Z\pi NN,V}^\mu\,,
\ea
and an axial part:
\ba
  J_{WNC,A}^\mu = J_{Z NN,A}^\mu + J_{Z\pi}^\mu + J_{Z\pi NN,A}^\mu\,.
\ea

\end{widetext}

\section{Nucleon resonances}\label{nucleon-resonances}

In this appendix we provide explicit expressions for the resonance excitation vertex ($Q NR$, electroweak vertex), the resonance decay vertex ($R\pi N$, strong vertex), the resonance propagators, the resonance form factors, and the resonance decay width.
The appendix is split into two parts: spin-3/2 and spin-1/2 resonances.

\subsection{Spin-3/2 resonances}

The electroweak vertex $Q NR_3$ ($Q$ stands for the boson, $N$ for the nucleon and $R_3$ for the spin-3/2 resonance) is given by the parametrization 
\ba
\Gamma_{QR_3 N}^{\beta\nu} = \left(\Gamma_{QR_3 N,V}^{\beta\nu} 
					+ \Gamma_{QR_3 N,A}^{\beta\nu} \right)\widetilde{\gamma}^5\,,
\ea
with $\widetilde{\gamma}^5 = \munit$ if the parity of the resonance is even, and $\widetilde{\gamma}^5 = \gamma^5$ if odd.
The vector part is given by
\begin{flalign}
\Gamma_{QR_3 N,V}^{\beta\nu} &= \Bigg[ 
	\frac{C_3^V}{M} (g^{\beta\nu}\Qslash - Q^\beta\gamma^\nu)&\non\\
	&+ \frac{C_4^V}{M^2} (g^{\beta\nu}Q\cdot K_{R} - Q^\beta K_{R}^\nu)&\non\\
	&+ \frac{C_5^V}{M^2} (g^{\beta\nu}Q\cdot P - Q^\beta P^\nu)
	+ C_6^Vg^{\beta\nu} \Bigg]\gamma^5\,,&
\end{flalign}
and the axial part is
\ba
\Gamma_{QR_3 N,A}^{\beta\nu} &=& 
	\frac{C_3^A}{M} (g^{\beta\nu}\Qslash - Q^\beta\gamma^\nu)\non\\
	&+& \frac{C_4^A}{M^2} (g^{\beta\nu}Q\cdot K_{R} - Q^\beta K_{R}^\nu)\non\\
	&+& C_5^A g^{\beta\nu} + \frac{C_6^A}{M^2} Q^{\beta}Q^{\nu}\,.
\ea
The form factors of the resonance, $C_i^{V,A}(Q^2)$, are described below. Here, $K_R^\mu$ stands for $K_s^\mu=P^\mu+Q^\mu$ or $K_u^\mu=P^\mu-K_\pi^\mu$.\\

{\sc $\Delta(1232)$ form factors.}
As mentioned in Sec.~\ref{Low-energy-model}, we have included the relative Olsson phases between ChPT and the $\Delta P$ contribution according to Ref.~\cite{Alvarez-Ruso16}. 
We do that by multiplying the vector form factors by the phase $\Psi_V$ and the axial ones by $\Psi_A$. 
In particular, we use the parameterization corresponding to `FIT~A'. 
To be consistent, we have used the same vector and axial form factors as in that fit (which in turn are from Ref.~\cite{Lalakulich06}), they are summarized below. 

The vector form factors are $C_6^V=0$ and
\ba
  C_3^V &=& \frac{2.13\ G_D^V}{1-Q^2/(4M_V^2)}\,,\non\\
  C_4^V &=& \frac{-1.51}{2.13}C_3^V\,,\\
  C_5^V &=& \frac{0.48\ G_D^V}{1-Q^2/(0.776M_V^2)}\,,  \non
\ea
with $G_D^V = (1-Q^2/M_V^2)^{-2}$ and $M_V=0.84$ GeV. 
The axial form factors are
$C_3^A=0$ and 
\ba
  C_5^A &=& C_5^A(0)(1-Q^2/M_{A,\Delta}^2)^{-2}\,,\non\\
  C_4^A &=& -C_5^A/4\,,\\
  C_6^A &=& C_5^A \frac{M^2}{m_\pi^2 - Q^2}\,,\non
\ea
with $C_5^A(0)=1.12$ and $M_{A,\Delta}=953.7$ MeV.

The general structure of the hadronic electroweak currents at quark level, along with isospin symmetry allow to relate the WNC, EM and CC form factors with each other (see, for instance, Section~3 in Ref.~\cite{Gonzalez-Jimenez13a} for details).
In this case, since the excitation of the Delta resonance is a purely isovector transition, the form factors that parameterize the nucleon-Delta transition vertex are the same for CC and EM interactions. Note that in the case of EM interactions only vector form factors should be considered.
On the other hand, the WNC form factors (denoted here as $\widetilde C_i^{V,A}$) are given by
\ba
  \widetilde C_i^V &=& (1-2\sin^2\theta_W)C_i^V\,,\non\\
  \widetilde C_i^A &=& C_i^A\,.
\ea

{\sc $D_{13}(1520)$ form factors.}
We describe the proton and neutron EM form factors by the parameterization given in Refs.~\cite{Lalakulich06}. For the proton one has
\ba
  C_3^{V,p} &=& \frac{-2.95\ G_D^V}{1-Q^2/(8.9M_V^2)}\,,\non\\
  C_4^{V,p} &=& \frac{-1.05}{2.95}C_3^{V,p}\,,\\
  C_5^{V,p} &=& 0.48\ G_D^V\,,\non
\ea
for the neutron 
\ba
  C_3^{V,n} &=& \frac{1.13\ G_D^V}{1-Q^2/(8.9M_V^2)}\,,\non\\
  C_4^{V,n} &=& \frac{-0.46}{1.13}C_3^{V,n}\,,\\
  C_5^{V,n} &=& 0.17\ G_D^V\,,\non
\ea
and $C_6^{V,p}=C_6^{V,n}=0$. We changed a global relative sign compared to Ref.~\cite{Lalakulich06} because of a different convention in the definition of the isovector form factors (see below).

The excitation of the $D_{13}$ resonance contains isovector and isoscalar contributions. 
Isospin symmetry allows one to relate proton and neutron form factors to isovector and isoscalar ones. 
One has
\ba
  C_i^{V,iv} &=& C_i^{V,p} - C_i^{V,n}\,,\non\\ 
  C_i^{V,is} &=& C_i^{V,p} + C_i^{V,n}\,,
\ea
where the labels $iv$ and $is$ stand for isovector and isoscalar. (In Ref.~\cite{Lalakulich06} the convention $C_i^{V,iv} = -(C_i^{V,p} - C_i^{V,n})$ was used.)
$C_i^{V,iv}$ are the form factors that enter in the CC interaction.

For the isovector axial form factors that enter in the CC interaction we use~\cite{Lalakulich06} $C_3^{A,iv}=C_4^{A,iv}=0$ and 
\ba
  C_5^{A,iv} &=& \frac{-2.1G_D^A}{1-Q^2/(3M_{A\Delta}^2)}\,,\non\\
  C_6^{A,iv} &=& C_5^{A,iv} \frac{M^2}{m_\pi^2 - Q^2}\,,
\ea
with $G_D^A = (1-Q^2/M_{A}^2)^{-2}$ and $M_A=1$ GeV. This is also used in Refs.~\cite{Hernandez13}.

The WNC form factors are given by
\ba
  \widetilde C_i^{V,p(n)} &=& \frac{1}{2}(1-4\sin^2\theta_W)C_i^{V,p(n)} \non\\
		      &-& \frac{1}{2}C_i^{V,n(p)} - \frac{1}{2}C_i^{V,s}\,,\\
  \widetilde C_i^{A,N} &=& (\pm)\frac{1}{2}C_i^A - \frac{1}{2}C_i^{A,s}\,,\non
\ea
where the $+$ ($-$) in the last equation stands for proton (neutron).
The strange contributions $C_i^{V(A),s}$ are unknown and, given the large uncertainties in the non-strange form factors, we fix them to zero.

The propagator of a spin-3/2 resonance $S_{R_{3},\alpha\beta}$ is described by 
\ba
S_{R_3,\alpha\beta} &=& \frac{-(\Kslash_{R} +M_{R})}{ K_{R}^2 - M_{R}^2 + iM_{R} \Gamma_{\text{{\tiny width}}}(W) }\non\\
  &\times& \Big( g_{\alpha\beta} -\frac{1}{3}\gamma_\alpha\gamma_\beta - \frac{2}{3 M_{R}^2}K_{R,\alpha} K_{R,\beta} \non\\
  &-& \frac{2}{3 M_{R}}(\gamma_\alpha K_{R,\beta} -K_{R,\alpha}\gamma_\beta )  \Big)\,,\label{eq:Sdelta}
\ea
which depends on the resonance-decay width $\Gamma_{\text{{\tiny width}}}(W)$. 

Finally, the $R_3\pi N$ vertex $\Gamma_{R_3\pi N}^\alpha$ reads
\ba
\Gamma_{R_3\pi N}^\alpha = \frac{\sqrt2 f_{\pi NR_3}}{m_\pi} K_\pi^\alpha\,\widetilde\gamma^5,
\ea
where the values of $f_{\pi NR_3}$ are given in Table~\ref{Res-properties}. \\

{\sc Decay width.}
We define the resonance-decay width as follows:
\ba
  \Gamma_{\text{{\tiny width}}}(W) = br\ \Gamma_{\text{{\tiny width}}}^{\pi N}(W)
	+ (1-br)\ \Gamma_{\text{{\tiny width}}}^{exp}\,,\non\\
\ea
where $W$ is the $\pi-N$ invariant mass and $br$ is the branching ratio for the pion-nucleon decay channel of the resonance.
$\Gamma_{\text{{\tiny width}}}^{exp}$ is the experimental value of the resonance-decay width (see Table~\ref{Res-properties}).
$\Gamma_{\text{{\tiny width}}}^{\pi N}(W)$ is the pion-nucleon resonance-decay width computed in the rest frame of the resonance~\cite{LeitnerThesis}
\ba
  \Gamma_{\text{{\tiny width}}}^{\pi N}(W) = 
    \frac{I_{iso}}{12\pi }\left(\frac{f_{\pi NR_3}}{m_\pi}\right)^2\frac{(k_{\pi}^*)^3}{W} (E^*_N\pm M).\non\\ \label{eq:R3-width}
\ea
The positive (negative) sign stands for even (odd) parity and $I_{iso}=1$ $(3)$ for isospin 3/2 (1/2). 
$k^*_\pi$ is the pion center of mass momentum given by Eq.~\ref{eq:Epi*} and $E_N=\sqrt{(k^*_\pi)^2+M^2}$. 

The values of the strong coupling constants $f_{\pi NR_3}$ are obtained from Eq.~\ref{eq:R3-width} by imposing $\Gamma_{\text{{\tiny width}}}(W=M_{R})=br\Gamma_{\text{{\tiny width}}}^{exp}$.
We use this value of $f_{\pi NR_3}$ both in the decay width (denominator of the current) and in the decay vertex (numerator of the current). Therefore, in the case $br<1$ our model will underestimate inclusive (no pion detected) cross-section data, since in the numerator we are only considering pion production through the process $R_3\rightarrow\pi N$, while in the denominator we are considering the full resonance-decay width, which may contain other decay channels such as $R_3\rightarrow\pi \Delta$, $R_3\rightarrow\eta N$, etc.

\subsection{Spin-1/2 resonances}

For spin-1/2 resonances the structure of the current and vertices is the same as that of the $NP$ and $CNP$ contributions.

The $QR_1N$ vertex is given by the sum of vector and axial contributions:
\ba
 \Gamma_{QR_1N}^\mu =  \left(\Gamma_{QR_1N,V}^\mu - \Gamma_{QR_1N,A}^\mu \right)\widetilde{\gamma}^5 \,.
\ea
The vector and axial vector parts are given by~\cite{Lalakulich06,Leitner09}  
\ba
 \Gamma_{QR_1N,V}^\mu &=& \frac{F_{1}}{\mu^2}(Q^\mu\Qslash - Q^2\gamma^\mu) 
	+ i\frac{F_2}{\mu}\sigma^{\mu\alpha}Q_\alpha\,,\non\\
 \Gamma_{QR_1N,A}^\mu &=& G_A\gamma^\mu\gamma^5 + \frac{G_P}{M}Q^\mu\gamma^5\,,
\ea
with $\mu=M_{R}+M$. The form factors are given below.\\

{\sc $P_{11}(1440)$ form factors.}
For the vector form factor of the proton and neutron we use the parameterization of Lalakulich et al.~\cite{Lalakulich06}.
For the proton
\ba
  F_1^p &=& \frac{-2.3\ G_D^V}{1-Q^2/(4.3M_V^2)}\,,\\
  F_2^p &=& -0.76\ G_D^V\{1-2.8\ln[1+Q^2/(1\text{GeV}^2)]\}\,.\non
\ea
Given the large uncertainties, in Ref.~\cite{Lalakulich06} they considered that the isoscalar contribution is negligible compare to the isovector one. Therefore, the neutron form factors are $F_{1,2}^n = -F_{1,2}^p$.
Due to the different convention for defining the isovector vector contribution, we have defined the proton form factors with a relative sign respect to those in Ref.~\cite{Lalakulich06}. Also, we have corrected for a relative wrong sign between $F_1^p$ and $F_2^p$ as it was pointed out in Ref.~\cite{Leitner09}.

The isovector vector form factors that enter in the CC interaction are defined as usual $F_{1,2}^{iv}=F_{1,2}^p-F_{1,2}^n$.
The isovector axial form factor is~\cite{Lalakulich06}
\ba
 G_A &=& \frac{0.51 G_D^A}{1-Q^2/(3M_A^2)} \,,
\ea
We changed the sign of this axial form factor respect to Ref.~\cite{Lalakulich06} so that the relative sign between vector and axial contributions match with the convention used in Refs.~\cite{Hernandez08, Leitner09}.
The pseudoscalar form factor is given by $G_P = \frac{\mu M}{m_\pi^2-Q^2}G_A$.

As in the case of the $D_{13}$ resonance, the WNC form factors are given by
\ba
  \widetilde F_i^{p,n} &=& \frac{1}{2}(1-4\sin^2\theta_W)F_i^{p,n} 
		    - \frac{1}{2}F_i^{n,p} - \frac{1}{2}F_i^{s}\,,\non\label{eq:v-wncFF}\\
  \widetilde G_A^{p,n} &=& \pm \frac{1}{2}G_A -\frac{1}{2} G_A^s\,,\label{eq:a-wncFF}
\ea
where the $+$ ($-$) in the last equation stands for proton (neutron). 
We fix the strange contributions to zero.\\

\begin{table}[htbp]
\centering
\begin{tabular}{c| c c c c c c c}
    & $I$ & $S$ & $P$ & $M_R$ & $\pi N$-$br$ & $\Gamma_{\text{{\tiny width}}}^{exp}$ & $f_{\pi NR}$\, \\
\hline
 $P_{33}$ & $3/2$ & $3/2$ &\ $+$ & 1232  & 100\% & 120 & 2.18  \\ 
 $D_{13}$ & $1/2$ & $3/2$ &\ $-$ & 1515  & 60\%  & 115 & 1.62  \\  
 $P_{11}$ & $1/2$ & $1/2$ &\ $+$ & 1430  & 65\%  & 350 & 0.391 \\ 
 $S_{11}$ & $1/2$ & $1/2$ &\ $-$ & 1535  & 45\%  & 150 & 0.16  \\
\hline
\end{tabular}
\caption{Properties of the resonances taken from Ref.~\cite{PDG08}. 
Masses and width are in MeV. $I$, $S$ and $P$ represent isospin, spin and parity, respectively. 
The coupling constants $f_{\pi NR}$ are computed using Eqs.~\ref{eq:R3-width} and \ref{eq:R1-width} (see text for details).}
\label{Res-properties}
\end{table}

{\sc $S_{11}(1535)$ form factors.}
We use the form factors of Ref.~\cite{Lalakulich06}. 
For the proton
\ba
  F_1^p &=& \frac{-2.0\ G_D^V}{1-Q^2/(1.2M_V^2)}\non\\
	&\times& \{ 1+7.2\ln[1-Q^2/(1\text{GeV}^2)]\}\,,\\
  F_2^p &=& -0.84\ G_D^V \{ 1+0.11\ln[1-Q^2/(1\text{GeV}^2)] \}\,.\non
\ea
Given the large uncertainties, in Ref.~\cite{Lalakulich06} the isoscalar contribution was considered to be negligible in comparison with the isovector one. Therefore, the neutron form factors are 
$F_{1,2}^n = -F_{1,2}^p$.
The isovector vector form factors are $F_{1,2}^{iv} = (F_{1,2}^p-F_{1,2}^n) = 2F_{1,2}^p$.
Due to a different convention in the definition of the isovector vector contribution, we have defined the proton form factors with a relative sign respect to those in Ref.~\cite{Lalakulich06}. 

The isovector axial form factors are~\cite{Lalakulich06}:
\ba
 G_A &=& \frac{-0.21 G_D^A}{(1-Q^2/(3M_A^2))} \,,\non\\
 G_P &=& \frac{(M_R-M)M}{m_\pi^2-Q^2}G_A \,,
\ea
with $M_A=1.05$ GeV.

The WNC form factors are given by Eqs.~\ref{eq:v-wncFF} and \ref{eq:a-wncFF}.

The propagator of a spin-1/2 resonance $S_{R_{1}}$ is
\ba
S_{R_1} = \frac{ \Kslash_{R} + M_{R} }{ K_{R}^2 - M_{R}^2 + iM_{R}\Gamma_{\text{{\tiny width}}}(W) }\,.\label{eq:SR1}
\ea
The resonance-decay width $\Gamma_{\text{{\tiny width}}}(W)$ is described below. 

Finally, the $R_1\pi N$ vertex $\Gamma_{R_1\pi N}$ is
\ba
\Gamma_{R_1\pi N}^\alpha = \frac{\sqrt2 f_{\pi NR_1}}{m_\pi} \Kslash_\pi\ \widetilde\gamma^5,
\ea
where the values of $f_{\pi NR_1}$ are given in Table~\ref{Res-properties}.\\

{\sc Decay width.} We use the same procedure as for spin-3/2 resonances. 
In this case, the resonance-decay width computed in the rest frame of the resonance reads~\cite{LeitnerThesis}:
\ba
\Gamma_{\text{{\tiny width}}}^{\pi N}(W) &=& 
    \frac{I_{iso}}{4\pi} \left(\frac{f_{\pi NR}}{m_\pi}\right)^2
    \frac{(W\pm M)^2}{W}\non\\
    &\times& (E^*_N\mp M) k^*_\pi\,.\label{eq:R1-width}
\ea
The upper (lower) sign corresponds to even (odd) parity.

\bibliographystyle{apsrev4-1}
\bibliography{bibliography}

\end{document}